\documentclass[journal]{IEEEtran}
\UseRawInputEncoding
\usepackage{array}
\usepackage{multirow}
\usepackage{colortbl}
\usepackage{booktabs}
\usepackage{makecell}
\usepackage{amssymb}
\usepackage{cite}
\usepackage{graphicx}
\usepackage{stfloats}
\usepackage{float}
\graphicspath{{./figures/}}
\usepackage{amsmath}
\newtheorem{theorem}{Theorem}

\usepackage{bm}
\usepackage{nomencl}
\usepackage[ruled,vlined]{algorithm2e}  
\usepackage{threeparttable}
\ifCLASSOPTIONcompsoc
\usepackage[caption=false,font=normalsize,labelfont=sf,textfont=sf]{subfig}
\else
\usepackage[caption=false,font=footnotesize]{subfig}
\fi
\usepackage{fancyhdr}
\ifCLASSINFOpdf
\else
\fi

\begin{document}
\title{An Ultra-Reliable Low-Latency Non-Binary Polar Coded SCMA Scheme}

\author{
	Shufeng Li,~\IEEEmembership{Member,~IEEE,}
        Mingyu Cai, Libiao Jin,
        Yao Sun,~\IEEEmembership{Senior Member,~IEEE,}
        Hongda Wu,~\IEEEmembership{Student Member,~IEEE,}
        and~Ping Wang,~\IEEEmembership{Fellow,~IEEE}
\thanks{This work was supported in part by the National Natural Science Foundation of China under Grant 61601414, and in part by the Fundamental Research Funds for the Central Universities under Grant CUC210B032.}
\thanks{Copyright (c) 2015 IEEE. Personal use of this material is permitted. However, permission to use this material for any other purposes must be obtained from the IEEE by sending a request to pubs-permissions@ieee.org.}
\thanks{Shufeng Li, Mingyu Cai and Libiao Jin are with the State Key Laboratory of Media Convergence and Communication, School of Information and Communication Engineering, Communication University of China, Beijing 100024, China (e-mail: lishufeng@cuc.edu.cn; cccmy1c@cuc.edu.cn; libiao@cuc.edu.cn).}
\thanks{Yao Sun is with James Watt School of Engineering, University of Glasgow, Glasgow, G12 8QQ, U.K. (e-mail: Yao.Sun@glasgow.ac.uk)}
\thanks{Hongda Wu and Ping Wang are with the Department of Electrical Engineering and Computer Science, Lassonde School of Engineering, York University, Toronto, ON M3J 1P3, Canada (e-mail: hwu1226@eecs.yorku.ca; pingw@yorku.ca)}
\thanks{Manuscript received XX XX, 20XX; revised XX XX, 20XX.}
}

\maketitle

\begin{abstract}
The joint transmission scheme of polar codes and sparse code multiple access (SCMA) has been regarded as a promising technology for future wireless communication systems. However, most of the existing polar-coded SCMA (PC-SCMA) systems suffer from high latency caused by the feedback iteration and list decoding. In addition, the error performance of PC-SCMA systems is unsatisfactory for ultra-reliable transmission. Inspired by the compelling benefits of non-binary polar codes, in this paper, we design a non-binary polar-coded SCMA (NB-PC-SCMA) system with a free order matching strategy to address the issues of delay and reliability. Specifically, we first formulate a joint factor graph for NB-PC-SCMA and propose a non-binary successive cancellation list (NB-SCL) and damping based joint iterative detection and decoding (NSD-JIDD) multiuser receiver to improve the BER and latency performance. Then, a lazy-search based NB-SCL (L-NB-SCL) decoding is proposed to reduce the computational complexity by simplifying the path search pattern of the list decoder. After that, we modify the update of user nodes for SCMA detection to improve the convergence error and finally propose the improved NSD-JIDD (ISD-JIDD) algorithm, which can avoid redundant operations by exploiting L-NB-SCL decoding. Simulation results show that the proposed NB-PC-SCMA system achieves better bit error rate (BER) performance and considerable latency gain when compared to its counterparts. In particular, the proposed ISD-JIDD can achieve similar BER performance of NSD-JIDD with less complexity.
\end{abstract}


\IEEEpeerreviewmaketitle

\section{Introduction}
\label{sec:1}

\IEEEPARstart{W}{ith} the rapid development of wireless communication networks, 
improving the spectrum efficiency with limited physical resources has become an 
urgent call for B5G/6G communication systems. One of the key technologies to 
solve this problem is the non-orthogonal multiple access (NOMA) scheme 
\cite{7263349}. Sparse code multiple access (SCMA), a potent code-domain NOMA 
scheme, is a new candidate for future communication systems attributed to the 
superior capacity of overload tolerance and resource reuse \cite{6666156}. At 
the receiver end, the message passing algorithm (MPA) is adopted to detect 
individual user symbols.

However, the SCMA system still fails to meet the throughput and reliability for future networks \cite{7973146}. In \cite{9432947}, the amalgamation with existing techniques, e.g., channel coding, was introduced to improve the performance of SCMA. So far, turbo coded SCMA (TC-SCMA) \cite{7248770,7848941,9399239} and low-density parity-check (LDPC) coded SCMA (LDPC-SCMA) \cite{7848813,8766141,8344383,8288397} systems have been widely investigated to obtain a coding gain. While these works conceived impressive joint detection and decoding (JDD) algorithms based on ``turbo principle'' for co-design systems, the above coding schemes are difficult to meet the ultra-reliable transmission for future communication systems and suffer a high computational complexity.
\begin{table*}[!t]
	\centering
	\renewcommand{\arraystretch}{1.3}
	\caption{Overview of existing literature on channel coded SCMA systems with 
	JDD schemes.}
	\label{table1}
	\resizebox{\textwidth}{!}{
		\begin{tabular}{c!{\vrule width0.9pt}c!{\vrule 
		width0.9pt}cccccccccccc!{\vrule width0.9pt}cc!{\vrule width0.9pt}cccc}
			\Xhline{1.2pt}
			\multirow{2}{*}{Contributions}& \multirow{2}{*}{This work}&
			\multicolumn{12}{c!{\vrule 
			width0.9pt}}{PC-SCMA}&\multicolumn{2}{c!{\vrule width0.9pt}}{ 
			TC-SCMA} &\multicolumn{4}{c}{LDPC-SCMA} \\ \cline{3-20}
			&  & \cite{8171004} & \cite{8463448} & \cite{9238845} & 
			\cite{ZHANG2020102283} & \cite{sym12101624} & \cite{8234623} & 
			\cite{8661315} & \cite{8543048} & \cite{9285274} & \cite{s20236740} 
			& \cite{9082596} & \cite{9386114} & \cite{7848941} & \cite{9399239} 
			& \cite{7848813} & \cite{8766141} & \cite{8344383} & \cite{8288397} 
			\\ \hline\hline
			
			\rowcolor[gray]{.85}Joint factor graph & \checkmark & \checkmark & 
			\checkmark & \checkmark  & \checkmark & \checkmark & & & & 
			\checkmark & \checkmark & & \checkmark & \checkmark & & \checkmark 
			& & \checkmark & \\
			
			soft-input soft-output & \checkmark & \checkmark & \checkmark & \checkmark  & 
			\checkmark & \checkmark & & & \checkmark & \checkmark & \checkmark 
			& \checkmark & \checkmark & \checkmark & \checkmark & \checkmark & 
			\checkmark & \checkmark & \checkmark \\
			
			\rowcolor[gray]{.85}Impact of user load & \checkmark & & & & & & 
			\checkmark & & & & & & \checkmark & & \checkmark & & & \checkmark & 
			\\
			
			BER improvement & \checkmark & \checkmark & \checkmark & 
			\checkmark  & \checkmark & & \checkmark & \checkmark & \checkmark & 
			\checkmark & \checkmark & \checkmark & \checkmark & \checkmark & 
			\checkmark & \checkmark & \checkmark & \checkmark & \checkmark \\
			
			\rowcolor[gray]{.85}Fading channels & \checkmark & & \checkmark & & 
			& & \checkmark & \checkmark & & \checkmark & \checkmark & 
			\checkmark & \checkmark & & \checkmark & & & & \\
			
			Complexity reduction & \checkmark & & & \checkmark  & \checkmark & 
			\checkmark & & \checkmark & & \checkmark & \checkmark & & 
			\checkmark & \checkmark & \checkmark & \checkmark & \checkmark & 
			\checkmark & \checkmark\\
			
			\rowcolor[gray]{.85}Latency reduction & \checkmark & & & & & & 
			\checkmark & & & \checkmark & & & & & \checkmark & & & \checkmark & 
			\\
			
			Early termination & \checkmark & & & & \checkmark & \checkmark & & 
			& & \checkmark & & & \checkmark & & \checkmark & \checkmark & & & \\
			
			\rowcolor[gray]{.85}Damping technique & \checkmark & & \checkmark & 
			\checkmark  & \checkmark & \checkmark & & & & & & & & \checkmark & 
			& \checkmark & & & \\
			
			Convergence analysis & \checkmark & & \checkmark & \checkmark & 
			\checkmark & & & & & & & \checkmark & \checkmark & \checkmark & 
			\checkmark & \checkmark & \checkmark & \checkmark & \checkmark\\
			
			\rowcolor[gray]{.85}Effect of channel estimation & & & & & & & & & 
			& & & \checkmark & \checkmark & & & & & \checkmark & \\
			
			Non-binary coding & \checkmark & & & & & & & & & & & & & & & & & 
			\checkmark & \checkmark\\
			
			\rowcolor[gray]{.85}Non-binary coding with FOMS & \checkmark & & & 
			& & & & & & & &  & & & & & & & \\
			\Xhline{1.2pt}
	\end{tabular}}
\end{table*}

As a standard code for control channels for 5G New Radio \cite{3GPP}, polar 
code proposed by E. Arikan in \cite{5075875} is the first coding scheme that 
can provably achieve the capacity of the binary input discrete memoryless 
channel with low complexity. In particular, polar codes can provide excellent 
error correction capability and higher spectral efficiency, which is 
competitive in the scenario of ultra-reliable low-latency communication (URLLC) 
\cite{8705373} for future wireless networks.

Benefiting from the above achievements, polar-coded SCMA (PC-SCMA) systems have 
been investigated in the literature. A typical JDD scheme directly combines 
soft-input soft-output polar decoder and SCMA detector 
\cite{8171004,8463448,9238845,ZHANG2020102283,sym12101624}. Specifically, the 
authors in \cite{8171004} proposed a JDD algorithm amalgamating MPA with belief 
propagation decoding to obtain performance gains. As a further development, a 
joint iteration detection and decoding (JIDD) receiver using a soft 
cancellation (SCAN) decoder was proposed in \cite{8463448} without inner 
iteration, which laid a foundation for JDD-based PC-SCMA receivers. In 
\cite{9238845,ZHANG2020102283,sym12101624}, some modified versions of JIDD were 
proposed to accelerate the convergence and reduce the computational complexity. 
However, none of these traditional soft-input soft-output schemes can break the BER performance limit of the SCAN decoder.

In contrast, an alternative JDD scheme amalgamates the soft-input hard-output polar decoder with the SCMA detector  \cite{8234623,8661315,8543048,9285274,s20236740,9082596,9386114}. To be more 
specific, a sequential user partition based JDD receiver was proposed in \cite{8234623,8661315} with limited receiver performance due to the feedback of hard outputs. The authors in \cite{8543048} first proposed a JDD scheme using a soft-input soft-output based successive cancellation (SC) decoder to achieve performance gains. Furthermore, a JIDD employing an SC-list (SCL) decoder was presented in \cite{9285274,s20236740}, which shows a better BER performance by designing the SCL decoder's extrinsic messages for turbo iteration. A joint channel estimation and detection scheme was also proposed for fading channels \cite{9082596,9386114}.

However, most of the soft-input hard-output schemes still lags behind the latency and BER target of URLLC. As a promising solution, non-binary polar codes (NB-PCs) can polarize discrete memoryless channels with arbitrary \emph{q}-ary alphabets for ultra-reliable transmission \cite{6303909}. Importantly, NB-PCs can save decoding latency by symbol-level operation instead of bit-level counterparts \cite{arXiv}, providing potential for URLLC application. Moreover, NB-PCs with different non-binary kernels over $GF(q)$ were investigated in \cite{7752615,8625284,9348796}, which shows significant BER gains over binary counterparts. 

The existing studies on coded SCMA systems employing JDD are summarized in 
Table \ref{table1}, which facilitates a comparison of this paper's 
contributions with other state-of-the-art research. As can be observed, these studies have mainly focused on BER improvement and complexity reduction, while there is a scarcity of literature on latency reduction, overload impact, early termination (ET), damping techniques and non-binary coding. Note that none of these work jointly designed coded SCMA systems with NB-PCs. The high decoding complexity of non-binary coding system increases the implementation difficulty and resource requirements during the hardware design. Besides, the inflexibility for the modulation scheme also hinders the wide application of non-binary coding, where the constellation points must be equal to the cardinality of the input alphabet.

Motivated by the impressive BER and latency performance of NB-PCs, we design a superior non-binary PC-SCMA (NB-PC-SCMA) architecture under a perfect channel state information condition\footnote{In this paper, the effect of channel estimation on the system performance is not considered, as shown in Table \ref{table1}. We assume that the channel state information is perfectly known for both transmitter and receiver.}. Overall, there are still challenges to achieve this target. First, most of the existing non-binary coding systems employ a modulation scheme that cannot be flexibly configured, i.e., constrained order matching strategy (COMS) \cite{8344383,8288397}, to facilitate the joint design, which solely considers a case that the finite field order is identical to the SCMA modulation order. COMS has an inherent weakness since it cannot trade-off system throughput and BER performance to fit the scenario. Second, if unconstrained order matching is considered, the receiver will yield the likelihood information conversion. Thus, the inner soft message exchange rules and the transmission reliability need to be resolved. Finally, although the latency and BER performance of the coded SCMA system are hopefully improved with NB-PCs, the receiver suffers an increased decoding complexity, which is unfriendly for hardware implementation. As a result, the receiver also requires a complexity reduction technique.
			
In this paper, we are engaged in tackling the above challenges to investigate the NB-PC-SCMA system. The main contributions of this paper are outlined as follows.
\begin{itemize}
	\item [1)] To the best of our knowledge, it is the first time to propose an NB-PC-SCMA scheme with free order matching strategy (FOMS). Specifically, we introduce symbol-to-bit conversion at the transmitter while the receiver applies a novel information exchange rule among the bits, field symbols, and SCMA codewords. With the aid of FOMS, the proposed system can freely select the field and modulation order configurations according to the requirements without the limitation of the NB-PC alphabet size.
	     
	\item [2)] Furthermore, we propose a non-binary SCL (NB-SCL) and damping 
	based JIDD (NSD-JIDD) algorithm for NB-PC-SCMA systems by combining the 
	factor graphs of the NB-PC decoder and the SCMA detector into a joint 
	factor graph (JFG). According to the connection in the JFG, the extrinsic 
	soft information is exchanged and is compressed by damping techniques to 
	improve error propagation. In addition, a cyclic redundancy check 
	(CRC)-based ET is applied to eliminate redundant iterations, which 
	facilitates clock cycle savings for the receiver.
	
	\item [3)] To reduce the complexity of the iterative receiver and enhance the convergence performance, the receiver constituents, i.e., the SCMA detector and the polar decoder, are improved. A lazy-search based NB-SCL (L-NB-SCL) decoding is proposed to avoid redundant path splitting in the decoding process. In addition, update operations that are less dominant for user nodes are removed in the SCMA detection process to make full use of the a priori information. Accordingly, the resultant improved NSD-JIDD (ISD-JIDD) algorithm can significantly decrease the computational complexity in the receiver.
	
\end{itemize}

The rest of this paper is organized as follows. Section \ref{sec:2} presents the FOMS and the proposed NB-PC-SCMA system model. Following this, a multiuser iterative receiver is designed in Section \ref{sec:3}, while its improved scheme is detailed in Section \ref{sec:4}. Section \ref{sec:5} discusses the performance of our proposed system over BER, complexity, and latency with numerical simulation results. Finally, the conclusions and future research prospects are drawn in Section \ref{sec:6}. Especially, the key abbreviations employed in this paper are summarized in Table \ref{table4} for ease of access.
\begin{table}[!t]
	\centering
	\renewcommand{\arraystretch}{1.3}
	\caption{Summary of abbreviations.}
	\label{table4}
\begin{tabular}{m{1cm}<{\centering}m{2.7cm}<{\centering}m{1cm}<{\centering}m{2.7cm}<{\centering}@{}}
		\toprule[1.2pt]
		Acronyms & Full Form & Acronyms & Full Form\\	
		\midrule[0.6pt]
		AWGN & additive white gaussian noise & NB-LDPC-SCMA & NB-LDPC coded SCMA\\
		BER & bit error rate & NB-PC & non-binary polar code\\
		COMS & constrained order matching strategy & NB-PC-SCMA & non-binary PC-SCMA\\
		CRC & cyclic redundancy check & NB-SCL & non-binary SCL\\
		ET & early termination & NSD-JIDD & NB-SCL and damping based JIDD\\
		FOMS & free order matching strategy & ISD-JIDD & improved NSD-JIDD\\
		IN & intermediate node & PC-SCMA & polar-coded SCMA\\
		JDD & joint detection and decoding & PN & polar node\\
		JFG & joint factor graph & RN & resource node\\
		JIDD & joint iteration detection and decoding & SC & successive cancellation\\ 
		LDPC & low-density parity-check & SCAN &soft cancellation\\
		LDPC-SCMA & LDPC coded SCMA & SCL & SC-list\\ 
		LLR & log-likelihood ratio & SCMA & sparse code multiple access\\  
		L-NB-SCL & lazy-search based NB-SCL & TC-SCMA & turbo coded SCMA\\ 
		MPA & message passing algorithm & UN & user nodes\\
		NB-LDPC & non-binary LDPC & URLLC & ultra-reliable low-latency communication\\ 	
		\bottomrule[1.2pt]	
	\end{tabular}
\end{table}

\section{System Model}
\label{sec:2}
\begin{figure}[!t]
	\centering
	\subfloat[COMS ($M = q$)]{\includegraphics[width=1.37in]{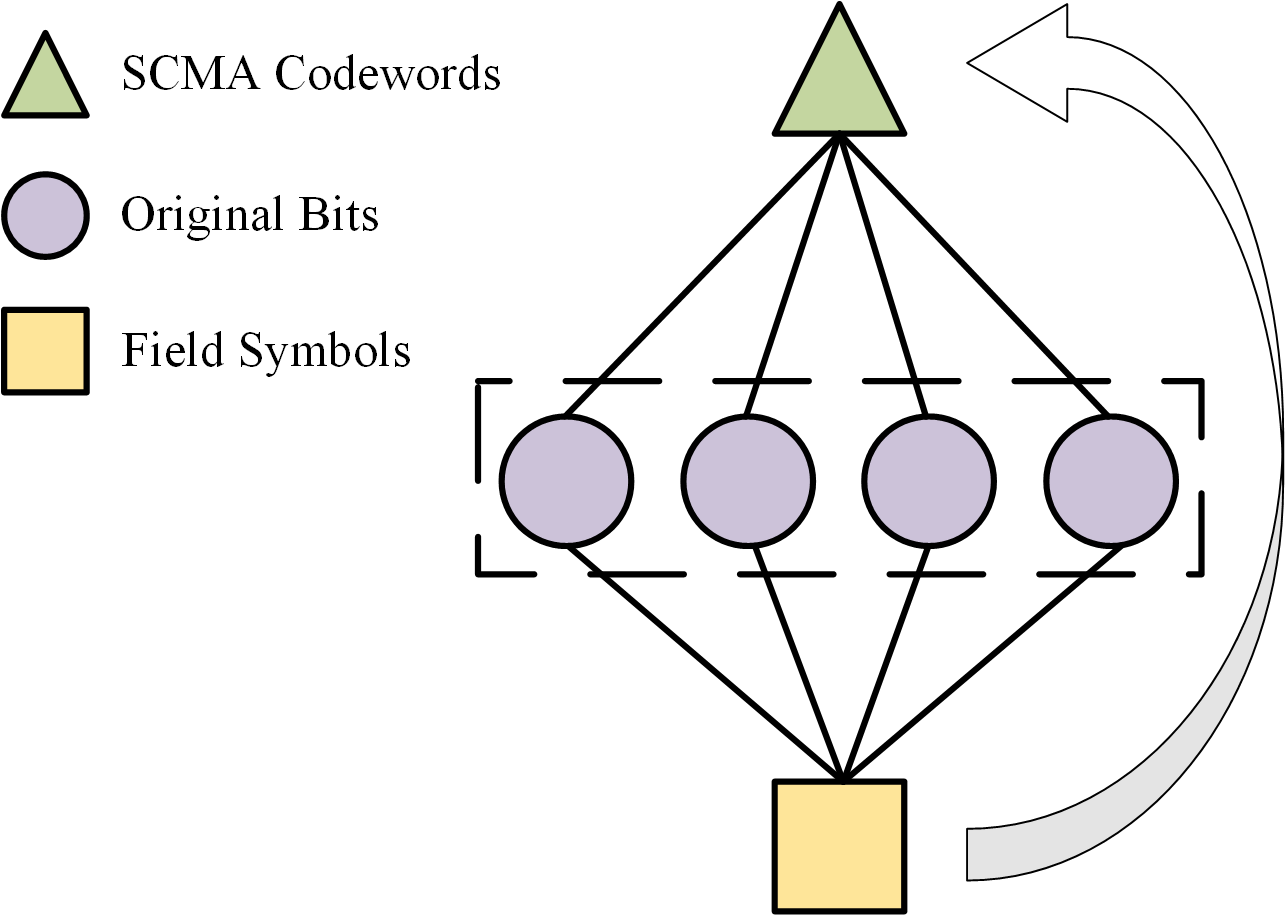}}
	\label{fig1-a}
	\\
	\subfloat[FOMS with $M < q$]{\includegraphics[width=1.67in]{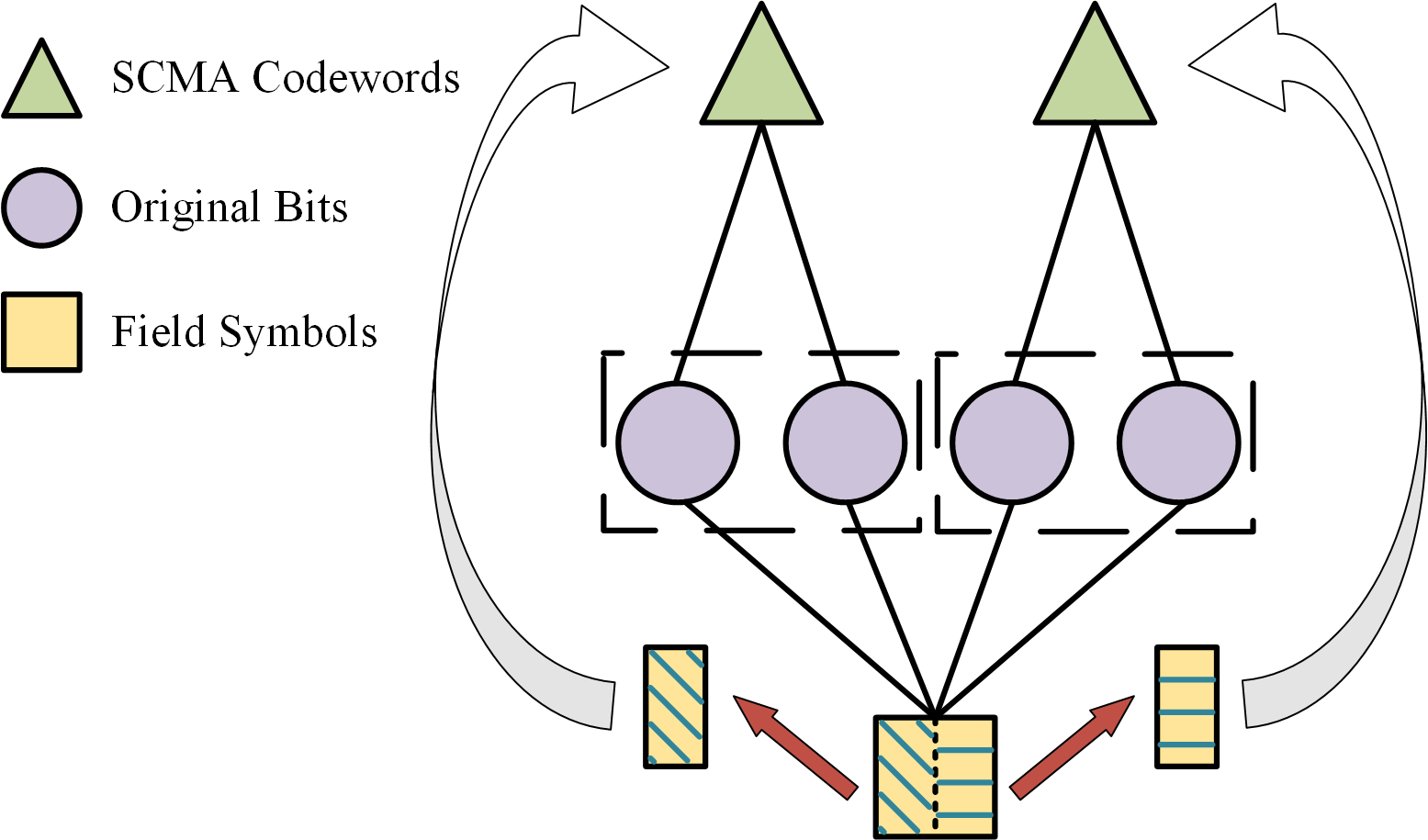}}
	\label{fig1-b}
	\hfil
	\subfloat[FOMS with $M > q$]{\includegraphics[width=1.45in]{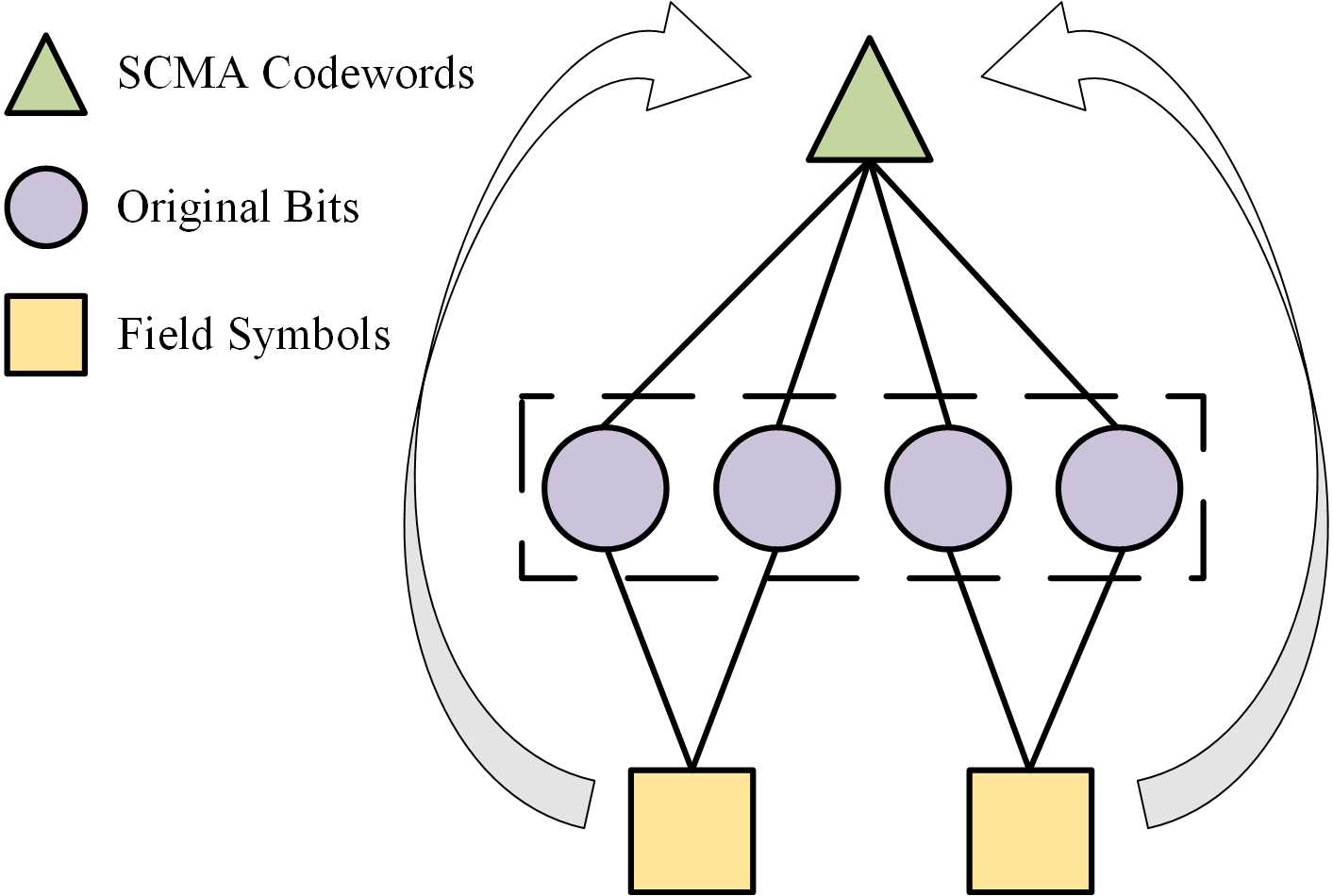}}
	\label{fig1-c}
	\caption{Different order matching strategies for non-binary coded SCMA.}
	\label{fig1}
\end{figure}
\begin{figure*}[!t]
	\centering
	\includegraphics[width=6.5in]{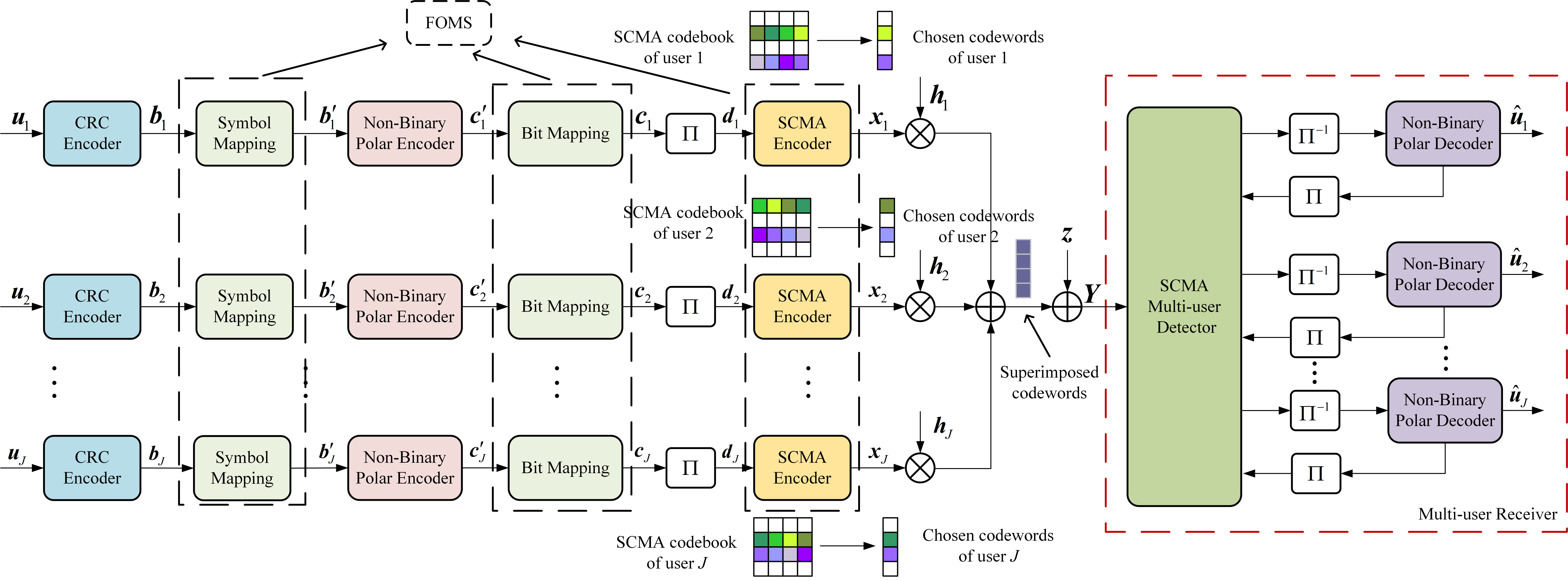}			
	\caption{FOMS based NB-PC-SCMA system.}
	\label{fig2}
\end{figure*}
\subsection{Free Order Matching Strategy}
\label{sec:2-1}
Assume that \emph{q}-ary codewords are converted into symbols by \emph{M}-point SCMA modulation at the transmitter. Fig. \ref{fig1} exemplifies the difference between COMS and FOMS under non-binary coded SCMA systems. For COMS of $q = 16$ shown in Fig. \ref{fig1}(a), every four bits are mapped into an SCMA codeword and Galois field symbol. Thus, SCMA codewords and field symbols are a group of one-to-one mapping. When $M = 4$ and $q = 16$, e.g., the case FOMS with $M < q$ in Fig. \ref{fig1}(b), a field symbol and an SCMA codeword are associated with 4 bits and 2 bits, respectively. In terms of field symbols, one symbol is mapped into two codewords. FOMS splits a field symbol into two sub-symbols using bit mapping. For $M = 16$ and $q = 4$, e.g., the FOMS with $M > q$ in Fig. \ref{fig1}(c), which is a typical multiple-to-one mapping, two field symbols are mapped into one codeword.

For COMS, only the system with $M = q$ is considered, which facilitates the modulation at the transmitter and the detection at the receiver. However, as the SCMA modulation order \emph{M} increases with \emph{q}, the receiver is more prone to misjudgment and has a worse convergence speed \cite{9260130}. Moreover, the system may suffer additional overhead due to the design of a large codebook.

Intuitively, COMS can be understood as a special case of FOMS with $M = q$. For FOMS, we introduce bit mapping at the transmitter, which makes adopting binary codewords for modulation feasible. Since the order matching is not limited, the universality is the visible advantage of FOMS. Importantly, the FOMS can trade-off the system throughput and reliability given the field order \textit{q} or SCMA modulation order \textit{M} according to the scene, which is demonstrated in Section \ref{sec:5-1}. The message exchange for FOMS-based receiver is depicted in Sections \ref{sec:3}.
\subsection{System Model for FOMS Based NB-PC-SCMA}
\label{sec:2-2}
The uplink NB-PC-SCMA system with FOMS is shown in Fig. \ref{fig2}. The data of \emph{J} users are multiplexed on \emph{K} orthogonal resources, giving an overload factor of $\lambda  = {J \mathord{\left/
{\vphantom {J K}} \right.\kern-\nulldelimiterspace} K}$. To be more specific, the \emph{A}-length information bits sent by user \emph{j} ($1 \le j \le J$) are denoted as ${{\bm{u}}_j} = \left[ {{u_{j,1}},{u_{j,2}}, \cdots ,{u_{j,A}}} \right]$, which is encoded as ${{\bm{b}}_j} = \left[ {{b_{j,1}},{b_{j,2}}, \cdots ,{b_{j,D}}} \right]$ by CRC encoder. The bits in ${\bm{b}_j}$ are converted to the field symbol over $GF\left( q \right)$ as the $D'$-length uncoded vector ${{\bm{b'\!\!}}_j} = \left[ {{{b'\!\!}_{j,1}},{{b'\!\!}_{j,2}}, \cdots ,{{b'\!\!}_{j,D'}}} \right]$, where field size $q = {2^p}$ and length $D' = {D \mathord{\left/{\vphantom {D p}}\right.\kern-\nulldelimiterspace} p}$. Then, ${{\bm{b'\!\!}}_j}$ are placed at $D'$ information symbol positions of sequence ${{\bm{a'\!\!}}_j} = \left[ {{{a'\!\!}_{j,1}},{{a'\!\!}_{j,2}}, \cdots ,{{a'\!\!}_{j,N'}}} \right]$, which are determined by Monte-Carlo simulation. The remaining positions are filled with the 0-valued frozen symbols. The resultant sequence ${{\bm{a'\!\!}}_j}$ is then encoded into ${{\bm{c'\!\!}}_j}$ containing $N' = {2^\omega }$ symbols by the non-binary polar encoder, which can be expressed as
\begin{equation}
	{{\bm{c'\!\!}}_j} = {{\bm{a'\!\!}}_j}{{\bm{G}}_2}^{ \otimes \omega },
	\label{1}
\end{equation}
where ${{\bm{G}}_2}^{ \otimes \omega }$ is the generator matrix of NB-PC and $ \otimes $ denotes the Kronecker product. According to \cite{5513568}, the kernel ${{\bm{G}}_2}$ can be achieved by extending the Arikan kernel to the Galois field, which is written as 
\begin{equation}
	{{\bm{G}}_2} = \left[ {\begin{array}{*{20}{c}}
			1&0\\
			\gamma &1
	\end{array}} \right],
	\label{2}
\end{equation}
where $\gamma  \in GF\left( q \right)\backslash 0$ represents the non-zero element over $GF\left( q \right)$.

Then, the output ${{\bm{c'\!\!}}_j}$ of the encoder is converted to the bit stream ${{\bm{c}}_j} = \left[ {{c_{j,1}},{c_{j,2}}, \cdots ,{c_{j,N}}} \right]$ by the bit mapper, where $N = pN'$. Here, we define the code rate as ${R_c} = {D \mathord{\left/{\vphantom {D N}} \right.\kern-\nulldelimiterspace} N} = {{D'} \mathord{\left/{\vphantom {{D'} {N'}}} \right.\kern-\nulldelimiterspace} {N'}}$.

\begin{figure}[!t]
	\centering
	\includegraphics[width=2in]{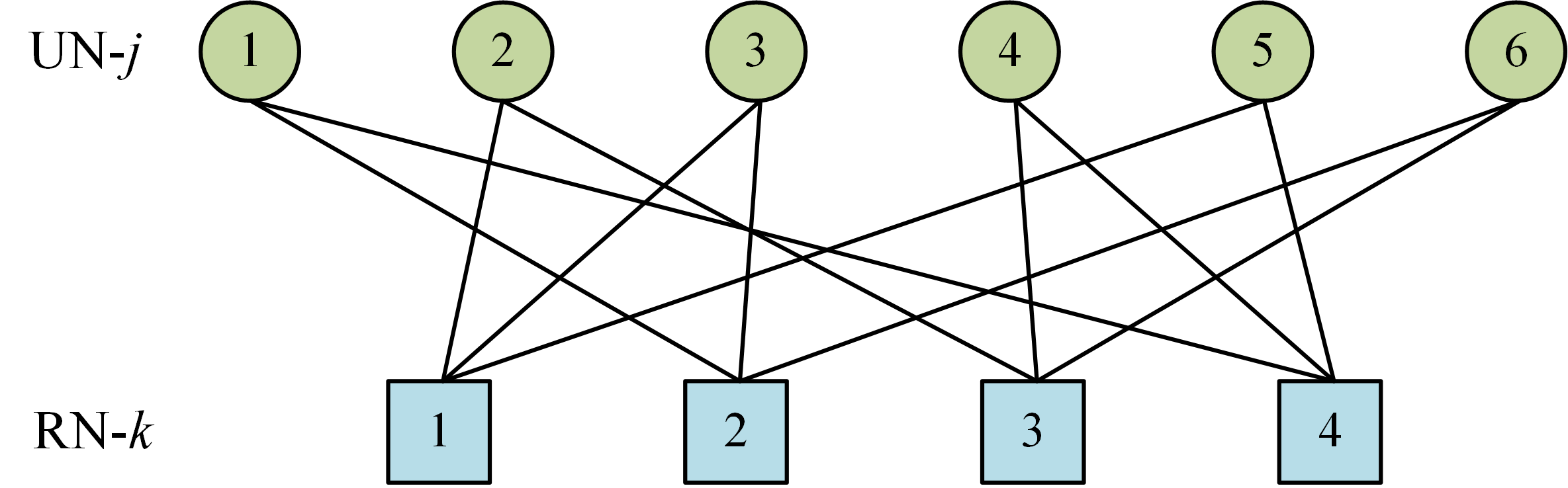}			
	\caption{SCMA factor graph for 6 users multiplexed over 4 resources.}
	\label{fig4}
\end{figure}

To mitigate interference caused by burst errors, ${{\bm{c}}_j}$ is interleaved by a random interleaver, which is expressed as ${{\bm{d}}_j} = \Pi ({{\bm{c}}_j})$, $1 \le j \le J$. ${{\bm{d}}_j}$ is then mapped to a \emph{K}-dimensional complex codeword ${{\bm{x}}_j} = \left[ {{{\bm{x}}_{j,1}},{{\bm{x}}_{j,2}}, \cdots ,{{\bm{x}}_{j,E}}} \right]$ by the SCMA encoder, where the \emph{e}-th ($1 \le e \le E$) codeword is a sparse column vector ${{\bm{x}}_{j,e}} = {\left[ {x_{j,e}^1,x_{j,e}^2, \cdots ,x_{j,e}^K} \right]^T}$. Supposing the modulation order (i.e., codebook cardinality) of SCMA is \emph{M}, the codeword symbol length is $E = {N \mathord{\left/{\vphantom {N R}} \right.\kern-\nulldelimiterspace} R}$, where $R = {\log _2}M$. The resource sharing structure of SCMA can be represented by a $K \times J$ indicator matrix. For example, a case with 4 resources and 6 users is denoted as
\begin{equation}
	{\bm{F}} = \left[ {\begin{array}{*{20}{c}}
			0&1&1&0&1&0\\
			1&0&1&0&0&1\\
			0&1&0&1&0&1\\
			1&0&0&1&1&0
	\end{array}} \right],
	\label{3}
\end{equation}
where the rows and columns of $\bm{F}$ represent the subcarrier and user layers, respectively. The \emph{j}-th user occupies the \emph{k}-th subcarrier if and only if the element ${{\bm{F}}_{kj}}$ in the \emph{k}-th row and \emph{j}-th column of $\bm{F}$ is 1. To be more specific, each user's data is assigned to ${d_u}$ (${d_u} \ll K$) resources, while ${d_r}$ (${d_r} \ll J$) users collide over the \emph{k}-th resource. The factor graph corresponding to the indicator matrix in (\ref{3}) is shown in Fig. \ref{fig4}. There are generally two types of nodes in a factor graph, i.e., user nodes (UNs) and resource nodes (RNs). Let UN-\emph{j} ($1 \le j \le J$) and RN-\emph{k} ($1 \le k \le K$) denote the \emph{j}-th UN and the \emph{k}-th RN, respectively.

Here, we assume that all users are time-synchronized and thus the \emph{e}-th received signal is the superposition of all users' signals, which can be expressed as
\begin{equation}
	{{\bm{y}}_e} = \sum\limits_{j = 1}^J {{\mathop{\bm diag}\nolimits} } \left( {{{\bm{h}}_{j,e}}} \right){{\bm{x}}_{j,e}} + {{\bm{z}}_e},
	\label{4}
\end{equation}
where ${{\bm{y}}_e} = {\left[ {y_e^1,y_e^2, \cdots ,y_e^K} \right]^T}$ is the received signal, ${{\bm{h}}_{j,e}}{\bm{ = }}\left[ {h_{j,e}^1,h_{j,e}^2, \cdots ,h_{j,e}^K} \right]$ is the channel gain vector and ${{\bm{z}}_e}$ is the $K \times 1$ additive Gaussian vector with element-wise 0 mean and covariance matrix ${N_0}{{\bm{I}}_K}$. Here, ${{\bm{I}}_K}$ denotes the $K \times K$ diagonal matrix. Note that $h_{j,e}^k$ in ${{\bm{h}}_{j,e}}$ denotes the channel gain of the \emph{e}-th transmitted codeword between resource \emph{k} and user \emph{j}, which is available to both transmitter and receiver with the assumption of perfect channel state information. Finally, all received signals in a transmission block can be represented as ${\bm{Y}} = \left[ {{{\bm{y}}_1},{{\bm{y}}_2}, \cdots ,{{\bm{y}}_E}} \right]$.

\section{Multiuser Iterative Receiver Design for NB-PC-SCMA Systems}
\label{sec:3}
In this section, the JFG and receiver for the proposed NB-PC-SCMA system are presented.
\subsection{Joint Factor Graph Model for FOMS Based NB-PC-SCMA}
\label{sec:3-1}
\begin{figure*}[!t]
	\centering
	\includegraphics[width=5in]{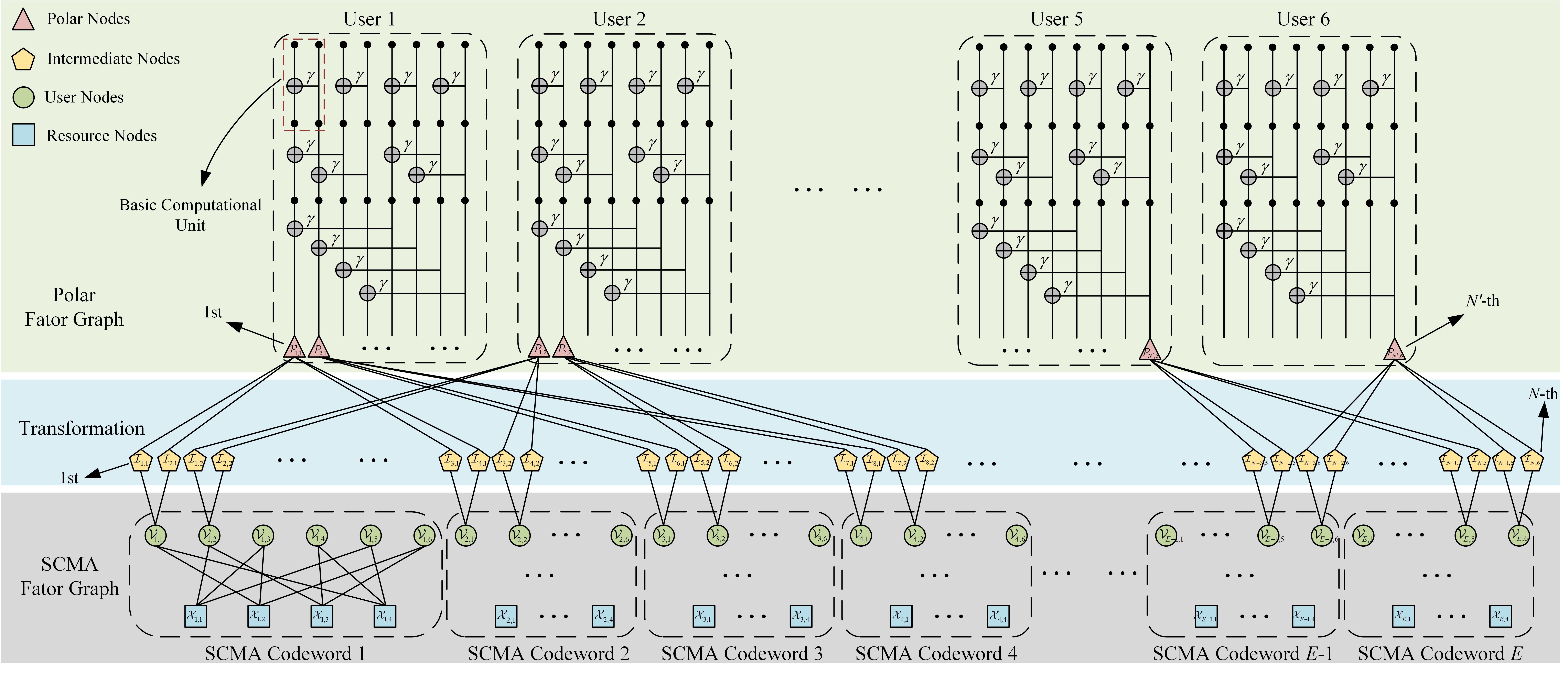}			
	\caption{The JFG for NB-PC-SCMA system.}
	\label{fig5}
\end{figure*}
To visualize our proposed receiver, we first introduce the JFG of NB-PC-SCMA, since the message exchange rules in the receiver solely rely on the connection of the JFG.  Moreover, the designed JFG equally reflects the impact of FOMS on the receiver, which corresponds to the characteristics regarding FOMS introduced in Section \ref{sec:2-1}.

Fig. \ref{fig5} exemplifies the JFG with $M = 4$, $q = 16$ and $N' = 8$, where 6 users and 4 resources are considered. Since FOMS allows $M \ne q$, the symbol-level MPA and NB-SCL are carried out over \emph{M}-ary and \emph{q}-ary, respectively. We build the mapping norm between the SCMA detector and the polar decoder, as shown in the transformation level in Fig. \ref{fig5}. Note that we term the node on the side of the polar decoder receiving a priori information as the polar node (PN) and the node at the transformation level introduced by FOMS as the intermediate node (IN). Considering \emph{J} users and \emph{E} SCMA codewords, JFG consists of \emph{J} polar factor graphs, \emph{E} SCMA factor graphs and $JN$ INs. Each polar factor graph contains $N'$ PNs. Let ${{\cal X}_{e,k}}$ and ${{\cal V}_{e,j}}$ denote RN-\emph{k} ($1 \le k \le K$) and UN-\emph{j} ($1 \le j \le J$) of the \emph{e}-th ($1 \le e \le E$) codeword, respectively, and let ${{\cal I}_{n,j}}$ ($1 \le n \le N$) and ${{\cal P}_{n'\!,j}}$ ($1 \le n' \le N'$) denote the \emph{n}-th IN and $n'$-th PN of the \emph{j}-th user, respectively.

Here, each UN and PN is associated with 2 INs and 4 INs, respectively, according to Fig. \ref{fig1}(b). Specifically, each ${{\cal V}_{e,j}}$ is connected to ${{\cal I}_{2e - 1,j}}$ and ${{\cal I}_{2e,j}}$, while each ${{\cal P}_{n'\!,j}}$ is connected to ${{\cal I}_{4n'\! - 3,j}}$, ${{\cal I}_{4n'\! - 2,j}}$, ${{\cal I}_{4n'\! - 1,j}}$, and ${{\cal I}_{4n'\!,j}}$. Therefore, we can imagine that ${{\cal P}_{n'\!,j}}$ is linked to ${{\cal V}_{2n'\! - 1,j}}$ and ${{\cal V}_{2n'\!,j}}$.

We integrate the polar factor graph and SCMA factor graph of different symbol levels into one JFG using IN, where messages are passed iteratively. Thus, the receiver for the proposed NB-PC-SCMA system only requires outer iterations, i.e., the loop of the overall operation. There is no inner iteration in the component SCMA detector or polar decoder. Note that we only give the JFG example for $M < q$. The model for $M > q$ is similar except for the connection pattern of IN. The result for $M > q$ can be easily found by referring to the interpretation for the FOMS shown in Fig. \ref{fig1}(c).

\subsection{NB-SCL and Damping Based Joint Iterative Detection and Decoding}
\label{sec:3-2}
\begin{figure}[!t]
	\centering
	\includegraphics[width=3.5in]{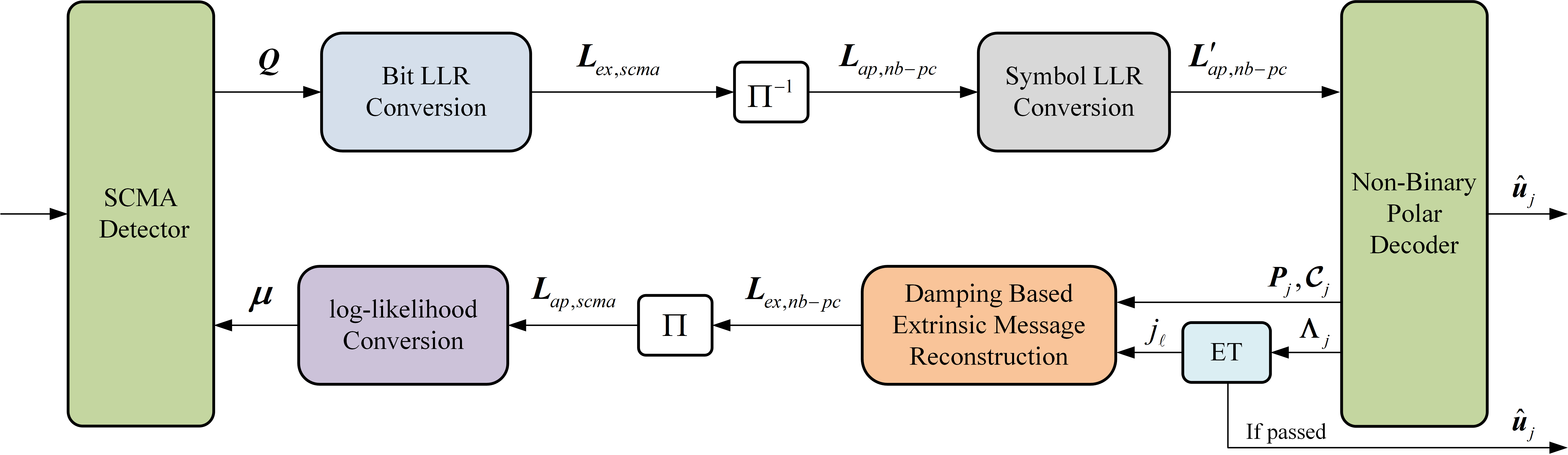}			
	\caption{Message passing in the NSD-JIDD receiver.}
	\label{fig6}
\end{figure}

In this section, we introduce the proposed multiuser receiver for NB-PC-SCMA systems, which jointly performs SCMA detection and polar decoding. Specifically, the NSD-JIDD message passing process is exemplified in Fig. \ref{fig6}, which is elaborated in Sections \ref{sec:3-2-1} to \ref{sec:3-2-4}.

\subsubsection{SCMA Detection}
\label{sec:3-2-1}
\ 
\newline
\indent 
The SCMA detection process can be interpreted as message passing on the factor graph. Therefore, the connection of the factor graph can be uniquely defined by a pair of sets ${\cal U}\left( j \right)$ and ${\cal R}\left( k \right)$ as follows,
\begin{subequations}
	\begin{equation}
		{\cal R}\left( k \right) = \left\{ {\left. j \right|{{\bm{F}}_{k,j}} = 1,1 \le j \le J} \right\},
		\tag{5.a}
		\label{5.a}
	\end{equation}
	\begin{equation}
		{\cal U}\left( j \right) = \left\{ {\left. k \right|{{\bm{F}}_{k,j}} = 1,1 \le k \le K} \right\}.
		\tag{5.b}
		\label{5.b}
	\end{equation}
	\label{5}
\end{subequations}
According to the max-log-MPA method \cite{8798841}, in the \emph{t}-th iteration, the messages are passed back and forth through the links in the factor graph and can be updated by (\ref{6}-\ref{7}). For the sake of convenience, we remove all subscripts \emph{e} indicating the codeword positions when
describing SCMA detector since the SCMA detection operation is identical for all codewords.

For UN-\emph{j} ($1 \le j \le J$), we have
\begin{equation}
	\begin{aligned}
		\xi _{j \to k}^{\left( t \right)}\left( {x_j^k = w_{j,k}^m} \right) = &{\cal N}\left( {{\mu ^{\left( {t - 1} \right)}}\left( {x_j^k = w_{j,k}^m} \right)} \right. \\
		&\left. { + \sum\limits_{i \in {\cal U}\left( j \right)\backslash k} {\xi _{i \to j}^{\left( {t - 1} \right)}\left( {x_j^i = w_{j,i}^m} \right)} } \right),
	\end{aligned}
	\label{6}
\end{equation}
while for RN-\emph{k} ($1 \le k \le K$), we have
\begin{equation}
	\begin{aligned}
		&\xi _{k \to j}^{\left( t \right)}\left( {x_j^k = w_{j,k}^m} \right)\\ = &\mathop {\max }\limits_{\scriptstyle x_i^k \in {{\cal W}_{i,k}},i \in {\cal R}\left( k \right)\backslash j\atop
		\scriptstyle x_j^k = w_{j,k}^m} \!\! \left\{ {\psi \left( {{{\bm{x}}_{[k]}}} \right) + \!\!\!\!\!\!\!\! \sum\limits_{x_i^k \in {{\bm{x}}_{[k]}},i \in {\cal R}\left( k \right)\backslash j}\!\!\!\! \!\!{\xi _{i \to k}^{\left( t \right)}\left( {x_i^k} \right)} } \right\},
	\end{aligned}
	\label{7}
\end{equation}
where $\xi _{k \to j}^{\left( t \right)}( {x_j^k = w_{j,k}^m})$ and $\xi _{j \to k}^{\left( t \right)}( {x_j^k = w_{j,k}^m} )$ are the messages sent from RN-\emph{k} to UN-\emph{j} and from UN-\emph{j} to RN-\emph{k} in the \emph{t}-th ($1 \le t \le T$) iteration given the codeword $x_j^k$, respectively. $w_{j,k}^m \in {{\cal W}_{j,k}}$ denotes the \emph{m}-th ($1 \le m \le M$) SCMA codeword of user \emph{j} transmitted by subcarrier \emph{k} in codebook ${\cal W}$. Here, ${\mu ^{\left( {t - 1} \right)}}( {x_j^k = w_{j,k}^m})$ denotes the a priori symbol log-likelihood information input in the $\left( {t - 1} \right)$-th iteration. ${\cal R}\left( k \right)\backslash j$ and ${\cal U}\left( j \right)\backslash k$ denote the set ${\cal R}\left( k \right)$ excluding \emph{j} and the set ${\cal U}\left( j \right)$ excluding \emph{k}, respectively. In (\ref{6}), ${\cal N}\left( \cdot \right)$ refers to the normalization function which ensures $\sum\limits_{m = 1}^M {\exp \left[ {\xi _{j \to k}^{\left( t \right)}( {x_j^k = w_{j,k}^m})} \right]}  = 1$. Assume that
\begin{equation}
	\begin{aligned}
		{\xi {_{j \to k}^{\left( t \right)} }}\!^ *\!\left( {x_j^k = w_{j,k}^m} \right) =& {\mu ^{\left( {t - 1} \right)}}\left( {x_j^k = w_{j,k}^m} \right)\\
		&+ \sum\limits_{i \in {\cal U}\left( j \right)\backslash k} {\xi _{i \to j}^{\left( {t - 1} \right)}\left( {x_j^i = w_{j,i}^m} \right)}.
	\end{aligned}
	\label{8}
\end{equation}
Then ${\cal N}\left( \cdot \right)$ can be further expressed as
\begin{equation}
	\begin{aligned}
		\xi _{j \to k}^{\left( t \right)}\left( {x_j^k = w_{j,k}^m} \right) = &{\xi {_{j \to k}^{\left( t \right)} }}\!^ *\!\left( {x_j^k = w_{j,k}^m} \right)\\
		&- \mathop {\max }\limits_{w_{j,k}^o \in {{\cal W}_{j,k}}} \left\{ {{\xi {_{j \to k}^{\left( t \right)} }}\!^ *\!\left( {x_j^k = w_{j,k}^o} \right)} \right\}.
	\end{aligned}
	\label{9}
\end{equation}
In addition, the vector function $\psi \left( {{{\bm{x}}_{[k]}}} \right)$ in (\ref{7}) can be defined as
\begin{equation}
	\psi \left( {{{\bm{x}}_{[k]}}} \right) =  - \frac{1}{{{N_0}}}{\left\| {{y^k} - \sum\limits_{i \in {\cal R}\left( k \right)} {h_i^kx_i^k} } \right\|^2},
	\label{10}
\end{equation}
where ${{\bm{x}}_{[k]}} = {\left[ {x_i^k} \right]_{i \in {\cal R}\left( k \right)}}$ is a vector comprising all symbols transmitted over the \emph{k}-th subcarrier.

Therefore, the soft information output by each user can be calculated as
\begin{equation}
	{Q^{\left( t \right)}}\left( {{{\bm{x}}_j} = {\bm{w}}_j^m} \right) = \sum\limits_{i \in {\cal U}\left( j \right)} {\xi _{i \to j}^{\left( t \right)}\left( {x_j^i = w_{j,i}^m} \right)},
	\label{11}
\end{equation}
where ${\bm{w}}_j^m \in {{\cal W}_j}$ ($1 \le m \le M$) represents the \emph{m}-th SCMA codeword of user \emph{j} in codebook ${\cal W}$. 
\subsubsection{Conversion and Calculation of LLRs under FOMS}
\label{sec:3-2-2}
\ 
\newline
\indent
The soft message output from the SCMA detector will be converted to bit level and further to field symbol level, which follows the JFG connection rigidly. Note that hereinafter we remove the superscript \emph{t} as all operations are performed within the same iteration.

To be more specific, the exchange messages are first converted to the available log-likelihood ratio (LLR) form for the polar decoder. The extrinsic bit LLR of the \emph{e}-th ($1 \le e \le E$) detected SCMA codeword for user \emph{j} can be expressed as
\begin{equation}
	{L_{ex,scma}}\left( {{{\hat d}_{j,\left( {e - 1} \right)R + r}}} \right) = \ln \frac{{\sum\limits_{s_{j,e}^i \in {\cal S}{_{j,e}},s_{j,e}^r = 0} {\exp \left( {Q\left( {{{\bm{x}}_{j,e}}} \right)} \right)} }}{{\sum\limits_{s_{j,e}^i \in {\cal S}{_{j,e}},s_{j,e}^r = 1} {\exp \left( {Q\left( {{{\bm{x}}_{j,e}}} \right)} \right)} }},
	\label{12}
\end{equation}
where ${\cal S}{_{j,e}} = \left\{ {s_{j,e}^1,s_{j,e}^2, \cdots ,s_{j,e}^R} \right\}$ ($1 \le r,i \le R$ and $i \ne r$) represents a bit set that can be mapped to SCMA codeword ${{\bm{x}}_{j,e}}$ by codebook ${{\cal W}_j}$ of user \emph{j} . In other words, each SCMA codeword message corresponds to \emph{R} bit messages, which conforms to the mapping relationship at the transmitter. Using the Jacobi approximation, we can further derive
\begin{equation}
	\begin{aligned}
		{L_{ex,scma}}\left( {{{\hat d}_{j,\left( {e - 1} \right)R + r}}} \right) =& \mathop {\max }\limits_{s_{j,e}^i \in {\cal S}{_{j,e}},s_{j,e}^r = 0} Q\left( {{{\bm{x}}_{j,e}}} \right)\\
		&- \mathop {\max }\limits_{s_{j,e}^i \in {\cal S}{_{j,e}},s_{j,e}^r = 1} Q\left( {{{\bm{x}}_{j,e}}} \right).
	\end{aligned}
	\label{13}
\end{equation}

After the extrinsic bit LLRs ${{\bm{L}}_{ex,scma}}( {{{{\bm{\hat d}}}_j}}) = [{{L_{ex,scma}}( {{{\hat d}_{j,1}}} ),{L_{ex,scma}}( {{{\hat d}_{j,2}}}), \cdots ,{L_{ex,scma}}( {{{\hat d}_{j,N}}})}]$ of all \emph{E} codewords for user \emph{j} are obtained, de-interleaving is performed to yield the a priori bit LLRs of the NB-SCL decoder, which can be expressed as
\begin{equation}
	{{\bm{L}}_{ap,nb - pc}}\left( {{{{\bm{\hat c}}}_j}} \right) = {\Pi ^{ - 1}}\left( {{{\bm{L}}_{ex,scma}}\left( {{{{\bm{\hat d}}}_j}} \right)} \right).
	\label{14}
\end{equation}

The de-interleaved bit LLRs will be converted into a priori symbol LLRs of NB-PCs and then input to the NB-SCL decoder. Before giving the conversion operation, we explain the symbol LLRs of NB-PCs. Unlike bit-level LLRs, the symbol LLRs with more than two likelihood information cannot be defined by a single ratio. We define the $n'$-th ($1 \le n' \le N'$) a priori symbol LLR vector of the non-binary polar decoder as

\begin{equation}
	\setcounter{equation}{15}
	\begin{aligned}
		{{\bm{L}}'_{ap,nb - pc}}({\hat c'_{j,n'}}) = [{L'_{ap,nb - pc}}&({\hat c'_{j,n'}} \!=\! 0),{L'_{ap,nb - pc}}({\hat c'_{j,n'}} \!=\! 1),\\
		&\cdots \!,{L'_{ap,nb - pc}}({\hat c'_{j,n'}} \!=\! {\alpha ^{q - 2}}){]^T}\!,
	\end{aligned}
	\label{15}
\end{equation}
where $\alpha $ is the primitive element of the finite field and a certain LLR is defined as
\begin{equation}
	{L'_{ap,nb - pc}}\left( {{{\hat c'}_{j,n'}} = \theta } \right) = \ln \frac{{\Pr \left( {{{\hat c'}_{j,n'}} = 0} \right)}}{{\Pr \left( {{{\hat c'}_{j,n'}} = \theta } \right)}},\theta  \in {\mathbb {F}_q},
	\label{16}
\end{equation}
where ${\mathbb {F}_q}$ denotes the set $\left\{ {0,1,\alpha ,{\alpha ^2}, \cdots ,{\alpha ^{q - 2}}} \right\}$ of all elements over $GF\left( q \right)$. 

\begin{theorem}
	For user \emph{j}, the $n'$-th NB-PC symbol LLR with estimate $\theta $ can be defined as
	\begin{equation}
		{L'_{ap,nb - pc}}\left( {{{\hat c'}_{j,n'}} = \theta } \right) = {{\cal X}_\theta }{{\cal L}_{j,n'}},
		\label{17}
	\end{equation}
	where ${{\cal X}_\theta } = \left[ {{v_1},{v_2}, \cdots ,{v_p}} \right]$ is a binary row vector satisfying the mapping relationship $\left\{ {f:{{\cal X}_\theta } \to \theta } \right\}$ over $GF\left( q \right)$ and ${{\cal L}_{j,n'}}$ is a column vector of a priori bit LLRs, which can be written as
	\begin{equation}
		\begin{aligned}
			{{\cal L}_{j,n'}} = [{L_{ap,nb - pc}}({\hat c_{j,\left( {n' - 1} \right)p + 1}}),&{L_{ap,nb - pc}}({\hat c_{j,\left( {n' - 1} \right)p + 2}}),\\
			&\cdots ,{L_{ap,nb - pc}}({\hat c_{j,n'p}}){]^T}.
		\end{aligned}
		\label{18}
	\end{equation}	
\end{theorem}

\emph{Proof:} Typically, the \emph{n}-th ($1 \le n \le N$) a priori bit LLR of user \emph{j} can be defined as
\begin{equation}
	{L_{ap,nb - pc}}\left( {{{\hat c}_{j,n}}} \right) = \ln \frac{{\Pr \left( {{{\hat c}_{j,n}} = 0} \right)}}{{\Pr \left( {{{\hat c}_{j,n}} = 1} \right)}}.
	\label{19}
\end{equation}

According to the connection of JFG described in Section \ref{sec:3-1}, we can conclude that ${{\cal P}_{n',j}}$ is associated with ${{\cal I}_{\left( {n' - 1} \right)p + 1,j}}$, ${{\cal I}_{\left( {n' - 1} \right)p + 2,j}}$, $\cdots$, ${{\cal I}_{n'p,j}}$, which implies that $\Pr( {{{\hat c'}_{j,n'}}})$ can be expressed as a joint probability mass function with respect to $\Pr( {{{\hat c}_{j,\left( {n' - 1} \right)p + i}}})$. When ${\hat c'_{j,n'}} = \theta $, the function can be written as
\begin{equation}
	\Pr \left( {{{\hat c'}_{j,n'}} = \theta } \right) = \prod\limits_{i = 1}^p {\Pr \left( {{{\hat c}_{j,\left( {n' - 1} \right)p + i}} = {v_i}} \right)} ,{v_i} \in {{\cal X}_\theta }.
	\label{20}
\end{equation}

Then, by substituting (\ref{20}) into (\ref{16}), we can derive
\begin{equation}
	\begin{aligned}
		{{L}'_{ap,nb - pc}}\left( {{{\hat c'}_{j,n'}} = \theta } \right)& = \sum\limits_{i = 1}^p {\ln } \frac{{\Pr \left( {{{\hat c}_{j,\left( {n' - 1} \right)p + i}} = 0} \right)}}{{\Pr \left( {{{\hat c}_{j,\left( {n' - 1} \right)p + i}} = {v_i}} \right)}}\\
		&= \!\!\sum\limits_{1 \le i \le p,{v_i} \ne 0} \!\!{{L_{ap,nb - pc}}\left( {{{\hat c}_{j,\left( {n' - 1} \right)p + i}}} \right)}.
	\end{aligned}
	\label{21}
\end{equation}
We can find that the result of (\ref{21}) is equivalent to (\ref{17}).$\hfill\blacksquare$

Note that ${L'_{ap,nb-pc}}\left( {{{\hat c'}_{j,n'}} = 0} \right)$ is always 
equal to 0 and thus does not require calculation via (\ref{17}). Finally, we 
can obtain the all a priori symbol LLRs of user \emph{j}, which can be 
expressed as ${{\bm{L}}'_{ap,nb - pc}}({{\bm{\hat c'}}_j}) = 
\left[{{\bm{L}}'_{ap,nb - pc}}({\hat c'_{j,1}}),{{\bm{L}}'_{ap,nb - pc}}({\hat 
c'_{j,2}}),\cdots ,{{\bm{L}}'_{ap,nb - pc}}({\hat c'_{j,N'}})\right]$

\subsubsection{N-SCl Decoding and Early Termination Mechanism}
\label{sec:3-2-3}
\ 
\newline
\indent
\begin{figure}[!t]
	\centering
	\includegraphics[width=3.5in]{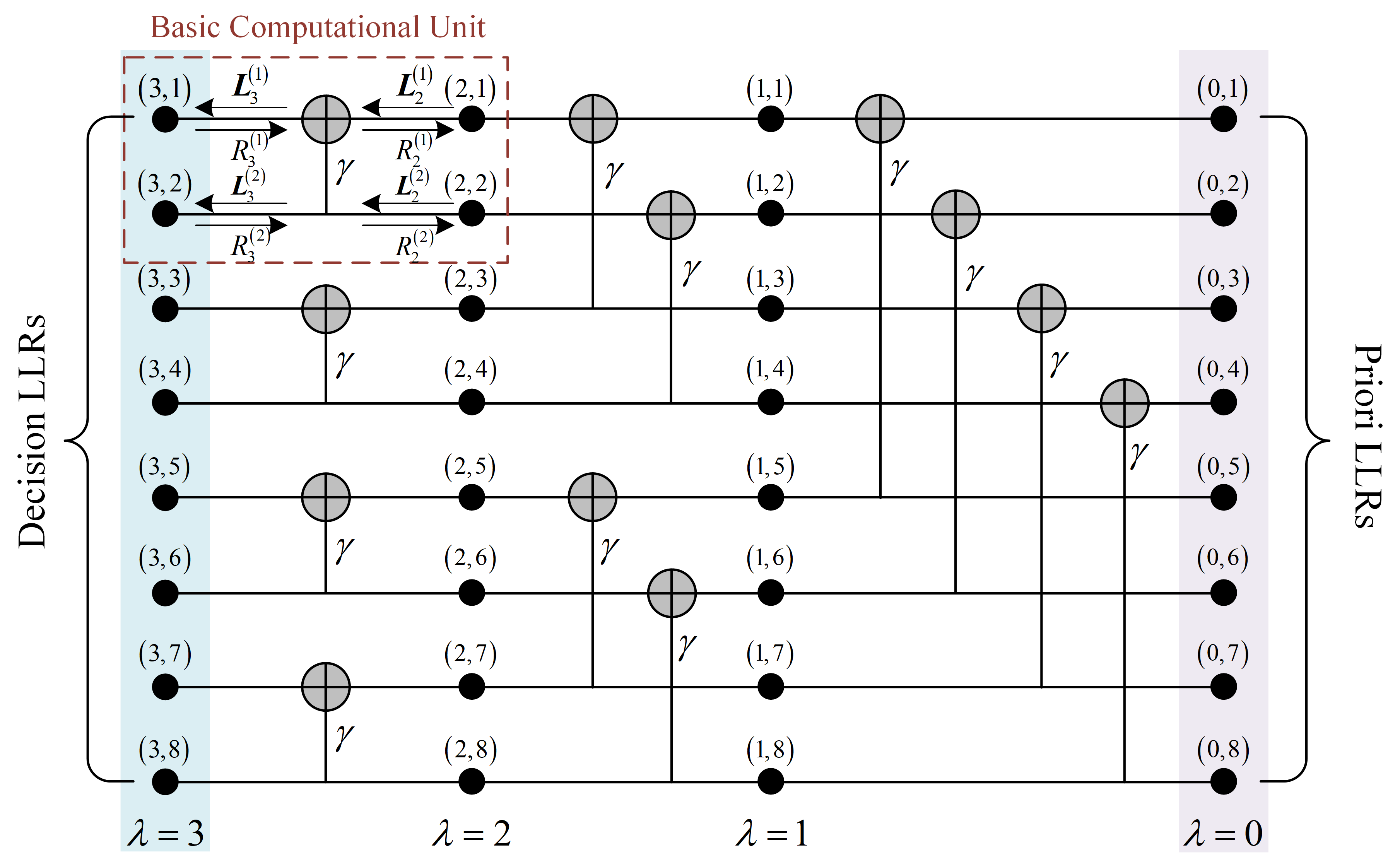}			
	\caption{The example of polar factor graph with $N' = 8$.}
	\label{fig7}
\end{figure}
Since the polar decoding process is equivalent for each user, we remove the subscript \emph{j} indicating users for brevity when describing polar decoder. After the transmitted soft information reaches the input of the non-binary polar decoder, i.e., the lower edge of the polar factor graph in the JFG shown in Fig. \ref{fig5}, the receiver will perform NB-SCL decoding.

To be more specific, as shown in Fig. \ref{fig7}, the non-binary polar factor graph with $N' = 8$ contains $\omega  + 1$ columns indexed by $\lambda $ and $N'$ rows indexed by $n'$. Each node $\left( {\lambda ,n'} \right)$ stores soft LLR information ${\bm{L}}_\lambda ^{(n')}$ passed to the left and hard symbol estimate $R_\lambda ^{(n')}$ passed to the right, where $1 \le n' \le N'$ and $0 \le \lambda  \le w$. Moreover, let ${{\bm{L}}_\lambda } = [{\bm{L}}_\lambda ^{(1)},{\bm{L}}_\lambda ^{(2)}, \cdots ,{\bm{L}}_\lambda ^{(N')}]$ and ${{\bm{R}}_\lambda } = [R_\lambda ^{(1)},R_\lambda ^{(2)}, \cdots ,R_\lambda ^{(N')}]$ denote all LLRs and estimates of the $\lambda $-th column, respectively.

To start with, the soft message ${{\bm{L}}_0}$ of each user's polar decoder is initialized as the received prior information ${{\bm{L}}'_{ap,nb-pc}}\left( {{\bm{\hat c'}}} \right)$, representing the rightmost input in Fig. \ref{fig7}. Then, the message will be passed through the basic computational unit in the factor graph, as shown in the red dashed box. We can obtain the update rules (\ref{23}-\ref{26}) for the ${{\bm{G}}_2}$-based basic computational unit by converting the probability-domain based recursive function \cite{arXiv} to the LLR form.
\begin{equation}
	\begin{aligned}
		L_\lambda ^{(n')}[\theta ] =& \mathop {\max }\limits_{^{\varphi  \in 
		{{\mathbb {F}}_q}}} \left\{ { - \sum\limits_{i = 1}^2 {L_{\lambda  - 
		1}^{(n')}[\varpi _{(0,\varphi )}^i]} } \right\}\\
		&- \mathop {\max }\limits_{^{\varphi  \in {{\mathbb {F}}_q}}} \left\{ { 
		- \sum\limits_{i = 1}^2 {L_{\lambda  - 1}^{(n' + {2^{_{w - \lambda 
		}}})}[\varpi _{(\theta ,\varphi )}^i]} } \right\},
		\label{23}		
    \end{aligned}
\end{equation}
\begin{equation}
	\begin{aligned}
		L_\lambda ^{(n' + {2^{_{w - \lambda }}})}[\theta ] =& \sum\limits_{i = 
		1}^2 {L_{\lambda  - 1}^{(n' + {2^{_{w - \lambda }}})}[\varpi 
		_{(R_\lambda ^{(n')},\theta )}^i]}\\ 
		&- \sum\limits_{i = 1}^2 {L_{\lambda  - 1}^{(n')}[\varpi _{(R_\lambda 
		^{(n')},0)}^i]},
		\label{24}	
	\end{aligned}
\end{equation}			
\begin{equation}
	R_{\lambda  - 1}^{\left( {n'} \right)} = R_\lambda ^{\left( {n'} \right)} + \gamma  \cdot R_\lambda ^{\left( {n' + {2^{_{w - \lambda }}}} \right)},
	\label{25}
\end{equation}
\begin{equation}
	R_{\lambda  - 1}^{\left( {n' + {2^{_{w - \lambda }}}} \right)} = R_\lambda ^{\left( {n' + {2^{_{w - \lambda }}}} \right)},
	\label{26}
\end{equation}
where $L_\lambda ^{(n')}[\theta ]$ ($\theta  \in {\mathbb {F}_q}$) in ${\bm{L}}_\lambda ^{(n')}$ denotes the LLR with the estimated value $\theta $, and $\gamma$ is a Galois field element of the kernel ${{\bm{G}}_2}$ defined in (\ref{2}). Besides, $\varpi _{\bm{\alpha}} ^i$ in (\ref{23}-\ref{24}) can be calculated as
\begin{equation}
	{{\bm{\varpi}} _{\bm{\alpha}} } = \left[ {\varpi _{\bm{\alpha}} ^1,\varpi _{\bm{\alpha}} ^2} \right] = {\bm{\alpha}}  \cdot {{{\bm{G}}}_2},
	\label{27}
\end{equation}
where ${\bm{\alpha }} \in {\mathbb {F}}_q^2$.

Note that $R_\lambda ^{(n')}$ in (\ref{25}-\ref{26}) actually represents the result of the re-encoding process for the decision symbols, i.e., the partial sum. In other words, (\ref{25}-\ref{26}) can be interpreted as mod-\textit{q} partial sum functions. The polar decoder will update them after decoding each symbol.

For the path metric of the NB-SCL decoder, we adjust the hardware-friendly function of the NB-PC path metric in \cite{9120673} to the compatible form with the LLR defined in (\ref{16}). Hence, for any path $\ell $ and level $n'$, the path metric $\rho _\ell ^{(n')}$ of the NB-SCL decoder can be calculated recursively as
\begin{equation}
	\rho _\ell ^{(n')} = \rho _\ell ^{(n' - 1)} + L_\omega ^{(n')}{\left[ \eta  \right]_{\left\langle \ell  \right\rangle }} - \mathop {\min }\limits_{\theta  \in {\mathbb {F}_q}} L_\omega ^{(n')}{\left[ \theta  \right]_{\left\langle \ell  \right\rangle }},
	\label{28}
\end{equation}
where $L_\omega ^{(n')}{\left[  \cdot  \right]_{\left\langle \ell  \right\rangle }}$ is the decision LLR for a given path $\ell $, representing the LLR at the decision layer shown in Fig. \ref{fig7}, and $\eta  \in {\mathbb {F}_q}$ is the $n'$-th estimate for the $\ell $-th path.

As a result, the decoded sequences with smaller path metrics can survive as candidates. Given a list size \emph{l} for the NB-SCL decoder, the \emph{l} most reliable paths and corresponding metrics of user \emph{j} are expressed as ${{\bm{\Lambda}} '_j} = \left[ {{{{\bm{\hat a}'}}_{{j_1}}},{{{\bm{\hat a}'}}_{{j_2}}}, \cdots ,{{{\bm{\hat a}'}}_{{j_l}}}} \right]$ and ${{\bm{P}}_j} = [\rho _{{j_1}}^{(N')},\rho _{{j_2}}^{(N')}, \cdots ,\rho _{{j_l}}^{(N')}]$, respectively, where ${{\bm{\hat a}'}_{{j_\ell }}}$ and $\rho _{{j_\ell }}^{(N')}$ ($1 \le \ell  \le l$) represent the $\ell $-th candidate symbol sequence and path metric, respectively. The codeword symbol sequences of the surviving paths are denoted as ${{\bm{{\cal C}}}'_j} = \left[ {{{{\bm{\hat c}'}}_{{j_1}}},{{{\bm{\hat c}'}}_{{j_2}}}, \cdots ,{{{\bm{\hat c}'}}_{{j_l}}}} \right]$, i.e., the hard estimate information ${{\bm{R}}_0}$ reaching the rightmost side, where ${{\bm{\hat c}'}_{{j_\ell }}}$ is the $\ell $-th codeword symbol sequence. After bit mapping, we can get \emph{l} candidate bit paths ${\bm{\Lambda} _j} = \left[ {{{{\bm{\hat a}}}_{{j_1}}},{{{\bm{\hat a}}}_{{j_2}}}, \cdots ,{{{\bm{\hat a}}}_{{j_l}}}} \right]$ and codeword bits ${\bm{{\cal C}}_j} = \left[ {{{{\bm{\hat c}}}_{{j_1}}},{{{\bm{\hat c}}}_{{j_2}}}, \cdots ,{{{\bm{\hat c}}}_{{j_l}}}} \right]$.

Assume that the $\ell $-th candidate path ${{\bm{\hat a}}_{{j_\ell }}}$ can pass the CRC and show the smallest path metric. Here, if all paths fail the CRC, ${{\bm{\hat a}}_{{j_\ell }}}$ only means the path with the smallest metric. Then the NB-SCL decoder outputs the parameter ${j_\ell }$ for the next stage and performs the CRC check for the next user. In the last iteration, the information bits corresponding to path ${{\bm{\hat a}}_{{j_\ell }}}$ are directly output as the estimated sequence ${{\bm{\hat u}}_j}$ of user \emph{j}. The receiver performs ET and directly outputs the estimated information bits ${{\bm{\hat u}}_1},{{\bm{\hat u}}_2}, \cdots ,{{\bm{\hat u}}_J}$ for all users if and only if the optimal path selected by each user passes the CRC.

\subsubsection{Damping based Extrinsic Message Reconstruction and Prior Information Update}
\label{sec:3-2-4}
\ 
\newline
\indent
Damping technique is an effective scheme to mitigate the error propagation problem and accelerate the convergence \cite{9234100}. In this paper, we introduce a damping factor $\varepsilon  \in (0,1]$ to compress the extrinsic message output by the polar decoder.

After NB-SCL decoding, the codeword bits ${{\bm{{\cal C}}}_j}$, the corresponding path metric ${{\bm{P}}_j}$, and the index ${j_\ell }$ of the selected path will be used for extrinsic message reconstruction. In this paper, we use the Bayes rule to calculate the likelihood information of codeword bits. The path metric $\rho _{{j_\ell }}^{(N')}$ of each candidate path is first normalized to
\begin{equation}
	{\delta _{{j_\ell }}} = {{\exp \left( { - \rho _{{j_\ell }}^{(N')}} \right)} \mathord{\left/
			{\vphantom {{\exp \left( { - \rho _{{j_\ell }}^{(N')}} \right)} {\sum\limits_{1 \le i \le l} {\exp \left( { - \rho _{{j_i}}^{(N')}} \right)} }}} \right.
			\kern-\nulldelimiterspace} {\sum\limits_{1 \le i \le l} {\exp \left( { - \rho _{{j_i}}^{(N')}} \right)} }}.
	\label{29}
\end{equation}

Then, the probability that the \emph{n}-th ($1 \le n \le N$) bit takes $\phi $ ($\phi  \in \left\{ {0,1} \right\}$) can be obtained by
\begin{equation}
	\Pr \left( {{{\hat c}_{j,n}} = \phi } \right) = \sum\limits_{1 \le \ell  \le l,{{{\bm{\hat c}}}_{{j_\ell }}} = \phi } {{\delta _{{j_\ell }}}}.
	\label{30}
\end{equation}

Therefore, the extrinsic bit LLR of the NB-SCL decoder can be written as
\begin{equation}
	{L_{ex,nb-pc}}\left( {{{\hat c}_{j,n}}} \right) = \left\{ {\begin{array}{*{20}{l}}
			{ - \infty }&{\Pr \left( {{{\hat c}_{j,n}} = 0} \right) = 0}\\
			{\ln \frac{{\Pr \left( {{{\hat c}_{j,n}} = 0} \right)}}{{\Pr \left( {{{\hat c}_{j,n}} = 1} \right)}}}&\begin{array}{l}
				\Pr \left( {{{\hat c}_{j,n}} = 0} \right) \ne 0 \\
				\&\Pr \left( {{{\hat c}_{j,n}} = 1} \right) \ne 0
			\end{array}\\
			{ + \infty }&{\Pr \left( {{{\hat c}_{j,n}} = 1} \right) = 0}
	\end{array}} \right..
	\label{31}
\end{equation}

We can correct the extrinsic bit LLR of user \emph{j} with the selected codeword bit path ${{\bm{\hat c}}_{{j_\ell }}}$, which can be expressed as
\begin{equation}
	{L_{ex,nb-pc}}\left( {{{\hat c}_{j,n}}} \right) = \left( {1 - 2{{\hat c}_{{j_\ell },n}}} \right)\left| {{L_{ex,nb-pc}}\left( {{{\hat c}_{j,n}}} \right)} \right|.
	\label{32}
\end{equation}

Before sent to the SCMA detector, the messages in the current iteration are directly damped by
\begin{equation}
	{\bm{L}}_{ex,nb - pc}^{(t)}\left( {{{{\bm{\hat c}}}_j}} \right) = \varepsilon {\bm{L}}_{ex,nb - pc}^{(t)}\left( {{{{\bm{\hat c}}}_j}} \right),
	\label{33}
\end{equation}
where ${\bm{L}}_{ex,nb - pc}^{(t)}$ denotes the extrinsic message of NB-PCs in the \emph{t}-th iteration.

Note that the extrinsic messages of SCMA are not moderated since the damping of the extrinsic messages output by MPA detectors and polar decoder lead to a similar a priori information behavior.

After the damping-aided extrinsic message reconstruction, we can obtain the extrinsic information of the polar decoder, which is then be interleaved as a priori bit LLR of the SCMA detector as
\begin{equation}
	{{\bm{L}}_{ap,scma}}({{\bm{{\hat d}}}_j}) = \Pi \left( {{{\bm{L}}_{ex,nb - pc}}\left( {{{{\bm{\hat c}}}_j}} \right)} \right).
	\label{34}
\end{equation}

Finally, the interleaved bit LLRs are remapped to the symbol log-likelihood of 
SCMA and are sent into the SCMA detector subsequently as the a priori 
information for the next iteration, which can be expressed as (\ref{36}) by 
employing the Jacobi approximation.
\begin{equation}
	\begin{aligned}
		\mu \left( {x_{j,e}^k} \right) = &\sum\limits_{s_{j,e}^r \in {\cal 
		S}{_{j,e}}} {\Big\{ \left( {1 - s_{j,e}^r} 
		\right){L_{ap,scma}}({{{\hat 
		d}_{j,\left( {e - 1} \right)R + r}}})}\\
		&- \max \left[ {0,{L_{ap,scma}}( {{{\hat d}_{j,\left( {e - 1} \right)R 
		+ r}}})} \right] \Big\}
		\label{36}		
	\end{aligned}
\end{equation}

Note that the calculation of $\mu \left( {x_{j,e}^k} \right)$ in (\ref{36}) is 
identical for $\forall k \in \left[ {1,K} \right]$, which only depends on the 
\emph{e}-th codeword mapped by the codebook ${{\cal W}_j}$. The proposed 
NSD-JIDD algorithm for the NB-PC-SCMA system is summarized in Algorithm 1.
\begin{algorithm}[t!] 
	\caption{NB-SCL and Damping Based Joint Iterative Detection and Decoding}
	\LinesNumbered 
	\KwIn{received signals $\bm{Y}$, maximum number of iterations \emph{T} and damping factor $\varepsilon $} 
	\KwOut{hard decisions of decoded bits ${{\bm{\hat u}}_1},{{\bm{\hat u}}_2}, \cdots ,{{\bm{\hat u}}_J}$} 
	\textbf{Initialization:} $\xi _{j \to k}^{( 0 )}( {x_j^k = w_{j,k}^m}) = 0$ and ${\mu ^{( 0 )}}( {x_j^k = w_{j,k}^m} ) = \log \frac{1}{M}$ for $\forall w_{j,k}^m \in {{\cal W}_{j,k}}$, $j = 1,2, \cdots ,J$ and $k = 1,2, \cdots ,K$\\
	\For{$t = 1,2, \cdots ,T$} 
	{
		Perform SCMA detection using (\ref{6}-\ref{7}) and (\ref{11})\;
		\For{$e = 1,2, \cdots ,E$} 
		{
			\For{$j = 1,2, \cdots ,J$}
			{
				Calculate extrinsic bit LLRs ${{\bm{L}}_{ex,scma}}( {{{{\bm{\hat d}}}_j}})$ of SCMA using (\ref{12});
			}		
		}
	    \For{$j = 1,2, \cdots ,J$}
	    {
	    	De-interleave ${{\bm{L}}_{ex,scma}}( {{{{\bm{\hat d}}}_j}})$ to obtain a priori bit LLRs ${{\bm{L}}_{ap,nb - pc}}( {{{{\bm{\hat c}}}_j}})$ of NB-PCs\;
	    	\For{$n' = 1,2, \cdots ,N'$}
	    	{
	    		Calculate the a priori symbol LLRs ${{\bm{L}}'_{ap,nb - pc}}( {{{{\bm{\hat c'}}}_j}})$ of NB-PCs for $\forall \theta  \in {{\mathbb F}_q}$ using (\ref{17}) to give the input of the NB-SCL decoder;
	        }
            Perform NB-SCL decoding for user \emph{j} using (\ref{23}-\ref{28}) to obtain ${\bm{\Lambda} _j}$, ${\bm{{\cal C}}_j}$ and ${{\bm{P}}_j}$\;
	    }
		\If {CRC passes}
		{
			Output the decoded sequences ${\bm{\hat u}_1},{{\bm{\hat u}}_2}, \cdots ,{{\bm{\hat u}}_J}$\;
		    Break; \ \tcp{Activate ET}
		}
	    \For{$j = 1,2, \cdots ,J$}
	    {
	    	Perform damping based extrinsic message reconstruction using (\ref{29}-\ref{33})\;
	    	Interleave ${{\bm{L}}_{ex,nb - pc}}\left( {{{{\bm{\hat c}}}_j}} \right)$ to get a priori bit LLRs ${{\bm{L}}_{ap,scma}}( {{{{\bm{\hat d}}}_j}})$ of SCMA\;	    
        }
        \For{$e = 1,2, \cdots ,E$} 
        {
        	\For{$j = 1,2, \cdots ,J$}
        	{
        		Calculate the a priori symbol log-likelihood of SCMA employing (\ref{36});
        	}		
        }		          
	}
    Get the decoded sequences ${\bm{\hat u}_1},{{\bm{\hat u}}_2}, \cdots ,{{\bm{\hat u}}_J}$\;
\end{algorithm}
\section{Improved Scheme of NSD-JIDD Algorithm}
\label{sec:4}
In this section, we conceive an improved implementation of NSD-JIDD, namely the ISD-JIDD algorithm. To be more specific, the L-NB-SCL decoding algorithm is proposed in Section \ref{sec:4-1} to simplify the path search pattern in NB-SCL decoding, leading to a complexity reduction. Then, in Section \ref{sec:4-2}, we modify the UN update for SCMA detection to mitigate the convergence errors during the iterations.
\begin{table*}[!t]
	\centering
	\renewcommand{\arraystretch}{1.3}
	\caption{The available range of lazy search rate with different code lengths and code rates.}
	\label{table2}
	\begin{tabular}{cccccccccccccc}
		\toprule[1pt]
		\multirow{2}{*}{\emph{q}}  & \multirow{2}{*}{$\beta$} & \multicolumn{3}{c}{$N' = 64$} & \multicolumn{3}{c}{$N' = 128$} & \multicolumn{3}{c}{$N' = 256$} & \multicolumn{3}{c}{$N' = 512$} \\
		\cmidrule(lr){3-3}\cmidrule(lr){4-4}\cmidrule(lr){5-5}\cmidrule(lr){6-6}\cmidrule(lr){7-7}\cmidrule(lr){8-8}\cmidrule(lr){9-9}\cmidrule(lr){10-10}\cmidrule(lr){11-11}\cmidrule(lr){12-12}\cmidrule(lr){13-13}\cmidrule(lr){14-14}
		&                       & ${R_c} = \frac{1}{3}$     & ${R_c} = \frac{1}{2}$    & ${R_c} = \frac{2}{3}$    & ${R_c} = \frac{1}{3}$     & ${R_c} = \frac{1}{2}$     & ${R_c} = \frac{2}{3}$    & ${R_c} = \frac{1}{3}$     & ${R_c} = \frac{1}{2}$     & ${R_c} = \frac{2}{3}$    & ${R_c} = \frac{1}{3}$     & ${R_c} = \frac{1}{2}$     & ${R_c} = \frac{2}{3}$    \\
		\cmidrule(lr){1-1}\cmidrule(lr){2-2}\cmidrule(lr){3-5}\cmidrule(lr){6-8}\cmidrule(lr){9-11}\cmidrule(lr){12-14}
		\multirow{2}{*}{4}  & $\beta _{\min }^ *$                  & 0.524   & 0.536  & 0.512  & 0.581   & 0.578   & 0.565  & 0.718   & 0.711   & 0.684  & 0.825   & 0.816   & 0.763  \\
		& $\beta _{\max }^ *$                & 0.667   & 0.656  & 0.628  & 0.767   & 0.734   & 0.729  & 0.859   & 0.836   & 0.801  & 0.942   & 0.918   & 0.850  \\
		\midrule[0.7pt]
		\multirow{2}{*}{16} & $\beta _{\min }^ *$                  & 0.571   & 0.563  & 0.558  & 0.721   & 0.719   & 0.706  & 0.871   & 0.859   & 0.825  & 0.924   & 0.910   & 0.889  \\
		& $\beta _{\max }^ *$                  & 0.762   & 0.750  & 0.721  & 0.837   & 0.828   & 0.812  & 0.941   & 0.930   & 0.895  & 0.959   & 0.941   & 0.936\\
		\bottomrule[1pt]
	\end{tabular}
\end{table*}
\begin{figure*}[!t]
	\centering
	\subfloat[Decoding tree of NB-SCL.]{\includegraphics[width=3.5in]{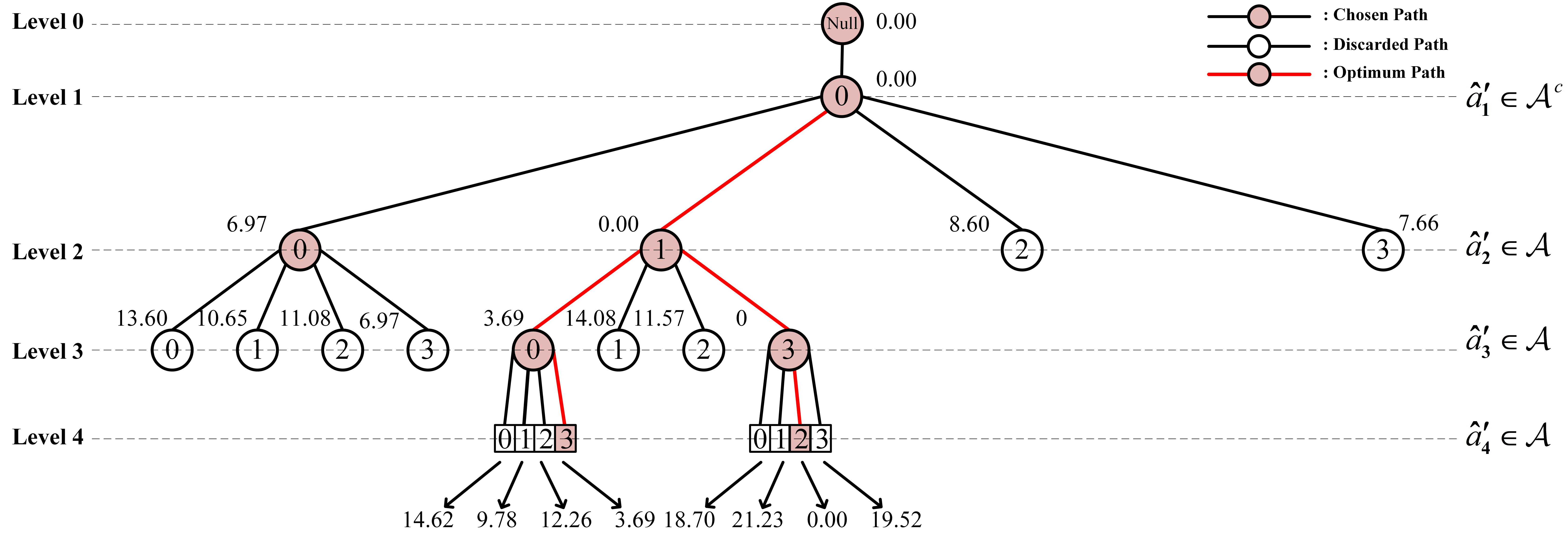}}
	\label{fig8-a}
	\hfil
	\subfloat[Decoding tree of L-NB-SCL.]{\includegraphics[width=3.5in]{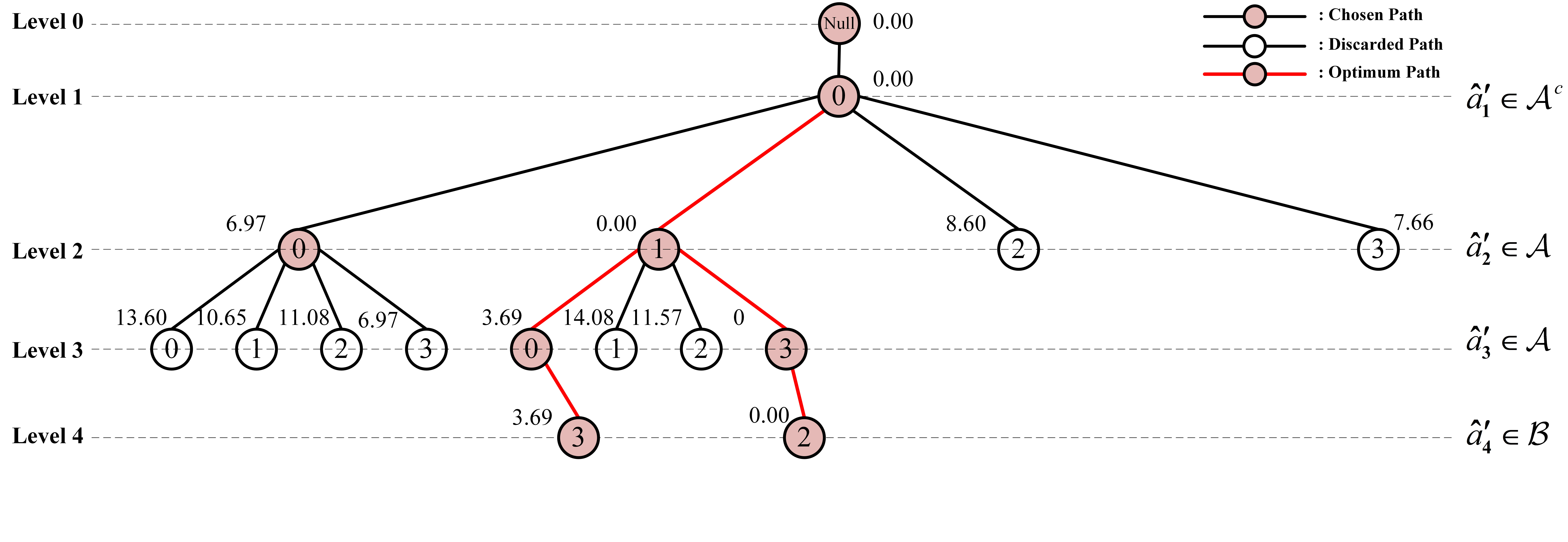}}
	\label{fig8-b}
	\caption{Example of a non-binary polarized decoding tree, where $N' = 4$ and $l = 2$.}
	\label{fig8}
\end{figure*}
\subsection{Lazy-Search Based NB-SCL Decoder}
\label{sec:4-1}
Since NB-SCL decoding adopts a global optimal search strategy over $GF(q)$, the decoder searches \emph{q} paths at each information position, leading to redundant computation. The Monte Carlo simulation shows that the symbols at certain positions are completely decoded correctly. Therefore, we consider these high-reliable positions can be decided directly during decoding. In other words, we introduce a local lazy search for NB-SCL decoding. The resultant decoding algorithm, i.e., L-NB-SCL decoding, only seeks paths at unreliable positions evaluated by Monte Carlo simulations. 

After the Monte Carlo simulation, let set ${\cal B}$ denote the position where the decision can be achieved directly, i.e., the high-reliable position. To be more specific, ${g_i}$ denotes the error rate of the \emph{i}-th ($1 \le i \le N'$) symbol in the Monte Carlo simulation. Then, we define the set of lazy positions as ${\cal B} = \left\{ {\left. i \right|{g_i} \le {g_{th}}} \right\}$ with cardinality $\chi  = \left| {\cal B} \right|$, where ${g_{th}}$ is the selected threshold with the constraint $0 \le {g_{th}} \le {10^{ - 3}}$ (assuming 10,000 Monte Carlo simulations are performed ). Naturally, set ${\cal B}$ is a subset of information position set $\,{\cal A}$. The lazy search rate is then defined as $\beta  = {\chi\left/\right.{D'}}$. In particular, the lazy search rates at ${g_{th}} = 0$ and ${g_{th}} = {10^{ - 3}}$ are denoted as $\beta _{\min }^ *$ and $\beta _{\max }^ *$, respectively. Therefore, we have the bound $\beta  \in \left[ {\beta _{\min }^ * ,\beta _{\max }^ * } \right]$.

Taking the $GF(4)$ and $GF(16)$ based NB-PC as an example, Table \ref{table2} gives the thresholds of $\beta$ for different code lengths $N'$ and code rates ${R_c}$, where Monte Carlo simulations are carried out at a signal-to-noise ratio of 2 dB. To obtain complexity reduction while maintaining performance, an appropriate $\beta$ is required to balance the trade-off between complexity and BER performance.

Different from the NB-SCL decoder, the L-NB-SCL decoder performs a lazy search, i.e., hard decision to reduce the split paths if $n' \in {\cal B}$. The hard decision function for estimate ${\hat a'_{\ell ,n'}}$ in the $\ell $-th path can be expressed as
\begin{equation}
	{\hat a'_{\ell ,n'}} = \mathop {\arg \min }\limits_{\theta  \in {{\mathbb {F}}_q}} L_\omega ^{({n'})}{\left[ \theta  \right]_{\left\langle \ell  \right\rangle }}.
	\label{37}
\end{equation}

Note that if $n' \in {\cal B}$, an update for the path metric is no longer required, i.e., $\rho _\ell ^{({n'})} = \rho _\ell ^{({n' - 1})}$, since the selected path complies with the SC decision, leading to a burden value of 0 in (\ref{28}).

The path search process in SCL decoding can be characterized by a decoding tree. Fig. \ref{fig8}(a) and  Fig. \ref{fig8}(b) show the $GF(4)$-based NB-SCL decoding tree and L-NB-SCL decoding tree, respectively. Typically, the $N'$-length $GF(q)$-based NB-PC decoding tree is a \emph{q}-ary tree with depth $N'$. Here, the topmost root node denotes a null state and the number adjacent to each node denotes the corresponding path metric. From Fig. \ref{fig8}, compared with NB-SCL decoding, L-NB-SCL decoding splits only one search path when the level $n' \in {\cal B}$. Generally, L-NB-SCL decoding can reach list saturation slower and avoid redundant global search.

Assume that the list is unfull with ${l_{pre}}$ paths surviving at the previous level. If the current stage belongs to ${\cal B}$, the search paths for L-NB-SCL and NB-SCL are ${l_{now}} = {l_{pre}}$ and ${l_{now}} = q{l_{pre}}$, respectively. Then, at the next information level belonging to ${{\cal B}^c}$ (the complement of the set ${\cal B}$), the search paths for both are ${l_{now}} = q{l_{pre}}$ and ${l_{now}} = {q^2}{l_{pre}}$, respectively. Thus, the calculation of $\left( {{q^2} - q} \right){l_{pre}}$ LLRs and path metrics are saved by L-NB-SCL decoding.

The L-NB-SCL decoding is summarized in Algorithm 2, where ${{\bm{R}}_{\lambda \left\langle \ell  \right\rangle }} = [ {{R{_\lambda ^{\left( 1 \right)}}}\!_{\left\langle \ell  \right\rangle },R{{_\lambda ^{\left( 2 \right)}}\!_{\left\langle \ell  \right\rangle }}, \cdots ,R{{_\lambda ^{( {N'})}}\!_{\left\langle \ell  \right\rangle }}}]$ ($0 \le \lambda  \le w$) denotes the right message of the $\lambda $-th column in the $\ell $-th factor graph. Note that for simplicity, the subscript \emph{j} representing user is omitted in Algorithm 2.
\begin{algorithm}[t!] 
	\caption{Lazy-Search Based NB-SCL Decoding}
	\LinesNumbered 
	\KwIn{maximum search width \emph{l}, information position set ${{\cal A}}$, high-reliable position set ${\cal B}$ and LLR initial value ${{\bm{L}}_0}$} 
	\KwOut{path metrics ${\bm{P}} = [ {\rho _1^{(N')},\rho _2^{( {N'} )}, \cdots ,\rho _l^{({N'})}} ]$, \emph{l} bit paths ${{\bm{\hat a}}_\ell } = \left[ {{{\hat a}_{\ell ,1}},{{\hat a}_{\ell ,2}}, \cdots ,{{\hat a}_{\ell ,N'}}} \right]$ and codeword bits ${{\bm{\hat c}}_\ell } = \left[ {{{\hat c}_{\ell ,1}},{{\hat c}_{\ell ,2}}, \cdots ,{{\hat c}_{\ell ,N'}}} \right]$}
	\textbf{Initialization:} ${\mathbb {L}} = \left\{ 1 \right\}$, $\rho _1^{\left( 0 \right)} = 0$\\
	{ 
		\For{$n' = 1,2, \cdots ,N'$} 
		{
			Update left message $L_\omega ^{({n'})}{[\eta ]_{\left\langle \ell  \right\rangle }}$ for $\forall \ell  \in {\mathbb {L}}$ and $\eta  \in {{\mathbb F}_q}$ using (\ref{23}-\ref{24})\;

		    \eIf{$n' \notin {\cal A}$} 
		    {
		    	${\hat a'_{\ell ,n'}} \leftarrow 0$ for $\forall \ell  \in {\mathbb {L}}$\; 
		    	Calculate the path metric $\rho _\ell ^{({n'})}$ for $\forall \ell  \in {\mathbb {L}}$ using (\ref{28})\; 
		    }
		    {
		    	\eIf{$n' \in {\cal B}$} 
		    	{ 
		    		Estimate ${\hat a'_{\ell ,n'}}$ for $\forall \ell  \in {\mathbb {L}}$ using (\ref{37})\;
		    		$\rho _\ell ^{( {n'})} \leftarrow \rho _\ell ^{( {n' - 1} )}$ for $\forall \ell  \in {\mathbb {L}}$\;
		    	}
		    	{
		    		Calculate temporary metric $\rho _{\ell ,\eta }^{temp}$ for $\forall \ell  \in {\mathbb {L}}$ and $\eta  \in {{\mathbb F}_q}$ using (\ref{28})\;
		    		Select $\min \left\{ {l,q\left| {\mathbb L} \right|} \right\}$ smallest $\rho _{\ell ,\eta }^{temp}$ and get survived symbol set ${{\mathbb Y}_\ell }$ of each  path\;
		    		
		    		\For{$\ell  \in {\mathbb {L}}$} 
		    		{
		    			\eIf{$\rho _{\ell ,\eta }^{temp}$ is seleted} 
		    			{ 
		    				$( {{{\hat a'}_{\ell ,n'}},\rho _\ell ^{( {n'})}}) \leftarrow ( {\eta ,\rho _{\ell ,\eta }^{temp}})$\;
		    				\If{$\left| {{{\mathbb Y}_\ell }} \right| > 1$}
		    				{
		    					\For{$\theta  \in {{\mathbb Y}_\ell }\backslash \eta $} 
		    					{
		    						Clone the path $\ell $ to a new path $\ell '$, ${\mathbb {L}} \leftarrow {\mathbb {L}} \cup \ell '$\;
		    						$( {{{\hat a'}_{\ell ',n'}},\rho _{\ell '}^{( {n'} )}} ) \leftarrow ( {\theta ,\rho _{\ell,\theta }^{temp}} )$\;
		    					}
		    			    }
		    				
		    			}
	    			    {
	    			    	Kill the path $\ell $ as ${\mathbb {L}} \leftarrow {\mathbb {L}} \backslash \ell $\;
	    			    }

%
		    		}	    	
		    	} 
		    } 
	        ${{R_0^{(n')}}_{\left\langle \ell  \right\rangle}} \leftarrow {\hat a_{\ell ,n'}}$\;
	        Update right message employing (\ref{25}-\ref{26})\;			
		}
	    \For{$\ell  = 1,2, \cdots ,l$} 
	    {
	    	${{\bm{\hat c'}}_\ell } \leftarrow {{\bm{R}}_{0\left\langle \ell  \right\rangle }}$\;
	    	Map ${{\bm{\hat a'}}_\ell }$ and ${{\bm{\hat c'}}_\ell }$ to ${{\bm{\hat a}}_\ell }$ and ${{\bm{\hat c}}_\ell }$, respectively 
	    }		
	}  
\end{algorithm}
\subsection{Modified MPA Detection}
\label{sec:4-2}
During the UN update of SCMA detection, the NSD-JIDD scheme employs the SCMA a priori information from the decoder and the information passed by RN in the prior iteration. However, as another constituent of the receiver, the SCMA detector outputs unreliable likelihood information at low ${{{E_b}} \mathord{\left/{\vphantom {{{E_b}} {{N_0}}}} \right.\kern-\nulldelimiterspace} {{N_0}}}$. To be specific, the information passed by RN commonly suffers a convergence error when ${{{E_b}} \mathord{\left/{\vphantom {{{E_b}} {{N_0}}}} \right.\kern-\nulldelimiterspace} {{N_0}}}$ is low, which leads to a BER degradation.

Explicitly, the dominant belief information passed by the UN should be the a priori information ${\mu ^{\left( {t - 1} \right)}}$ in (\ref{6}) since the latest a priori information from the non-binary polar decoder provides high reliability. By contrast, the passed information $\xi _{i \to j}^{\left( {t - 1} \right)}$ in the prior iteration is not an instant message and performs weakly at low ${{{E_b}} \mathord{\left/{\vphantom {{{E_b}} {{N_0}}}} \right.\kern-\nulldelimiterspace} {{N_0}}}$. After $\xi _{i \to j}^{\left( {t - 1} \right)}$ is employed to update ${{\bm{L}}_{ex,scma}}$ and ${{\bm{L}}_{ap,nb - pc}}$, the information ${\mu ^{\left( {t - 1} \right)}}$ will contain the main messages in $\xi _{i \to j}^{\left( {t - 1} \right)}$ and will be updated again.

As such, we modify the MPA detection in NSD-JIDD by removing the second term $\xi _{i \to j}^{\left( {t - 1} \right)}$ in (\ref{6}). In addition, the a priori information ${\mu ^{\left( {t - 1} \right)}}$ is strictly normalized since ${{\bm{L}}_{ex,nb - pc}}$ is constructed by the Bayes rule. Then, the ${\cal N}\left(  \cdot \right)$ operation can be discarded and thus save calculations. The modified UN update can be expressed as
\begin{equation}
	\xi _{j \to k}^{\left( t \right)}\left( {x_j^k = w_{j,k}^m} \right) = {\mu ^{\left( {t - 1} \right)}}\left( {x_j^k = w_{j,k}^m} \right).
	\label{38}
\end{equation}

Then, we can obtain the modified RN update rule by substituting (\ref{38}) into (\ref{7}) as follows,
\begin{equation}
	\begin{aligned}
		&\xi _{k \to j}^{\left( t \right)}\left( {x_j^k = w_{j,k}^m} \right)\\ = &\mathop {\max }\limits_{\scriptstyle x_i^k \in {{\cal W}_{i,k}},i \in {\cal R}\left( k \right)\backslash j\atop
			\scriptstyle x_j^k = w_{j,k}^m} \!\! \left\{ {\psi \left( {{{\bm{x}}_{[k]}}} \right) + \!\!\!\!\!\!\!\!\!\!\!\! \sum\limits_{x_i^k \in {{\bm{x}}_{[k]}},i \in {\cal R}\left( k \right)\backslash j}\!\!\!\! \!\!\!\!\!\!{{\mu ^{\left( {t - 1} \right)}}\left( {x_i^k} \right)} } \right\}.
	\end{aligned}
	\label{39}
\end{equation}

Therefore, the UN update in the modified MPA can be merged into the a priori information update. In this way, the ISD-JIDD does not comprise the UN update process, where the MPA detection stage can be represented by (\ref{39}).

\begin{figure}[!t]
	\centering
	\subfloat[NSD-JIDD.]{\includegraphics[width=3in]{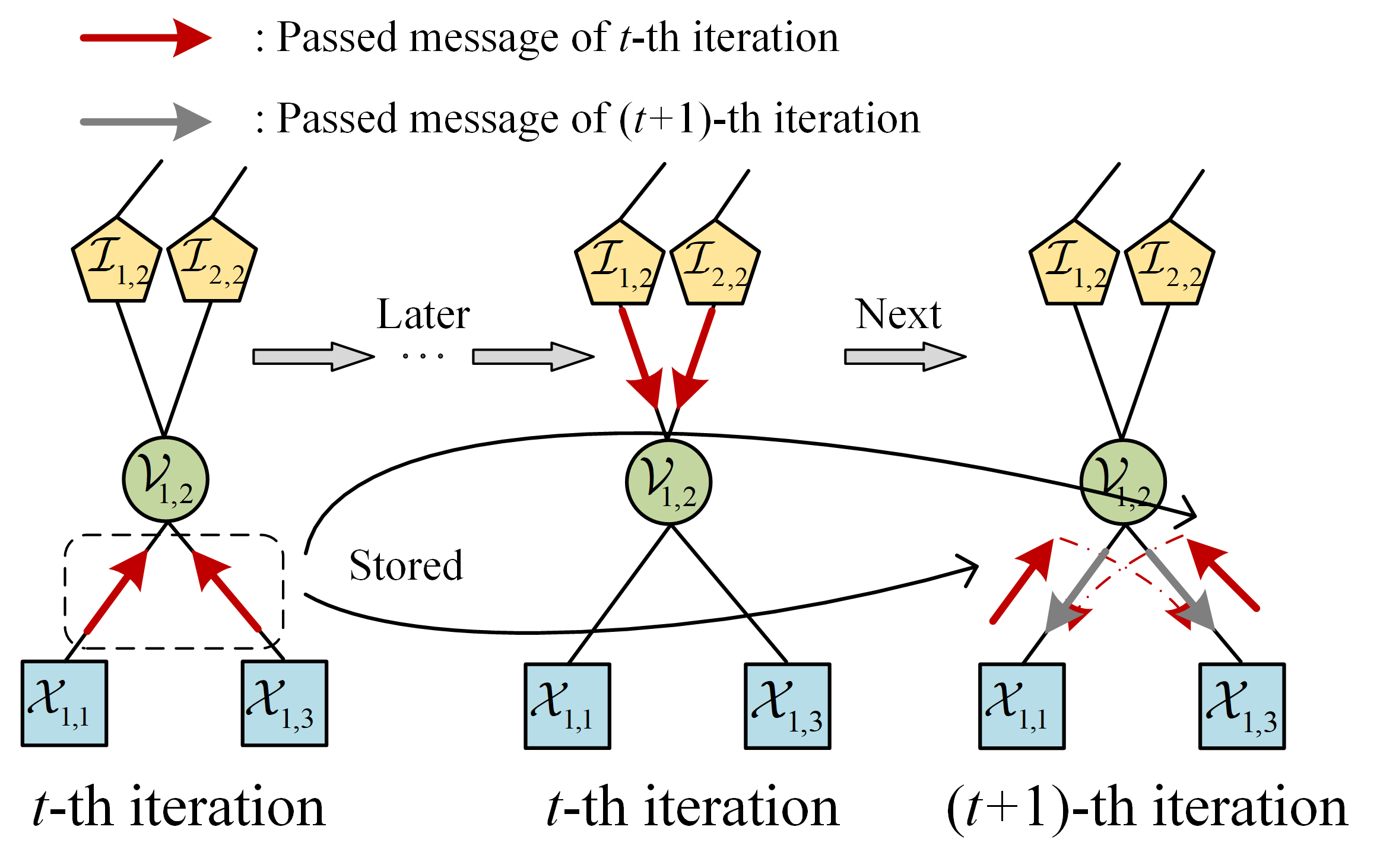}}
	\label{fig9-a}
	\subfloat[ISD-JIDD.]{\includegraphics[width=3in]{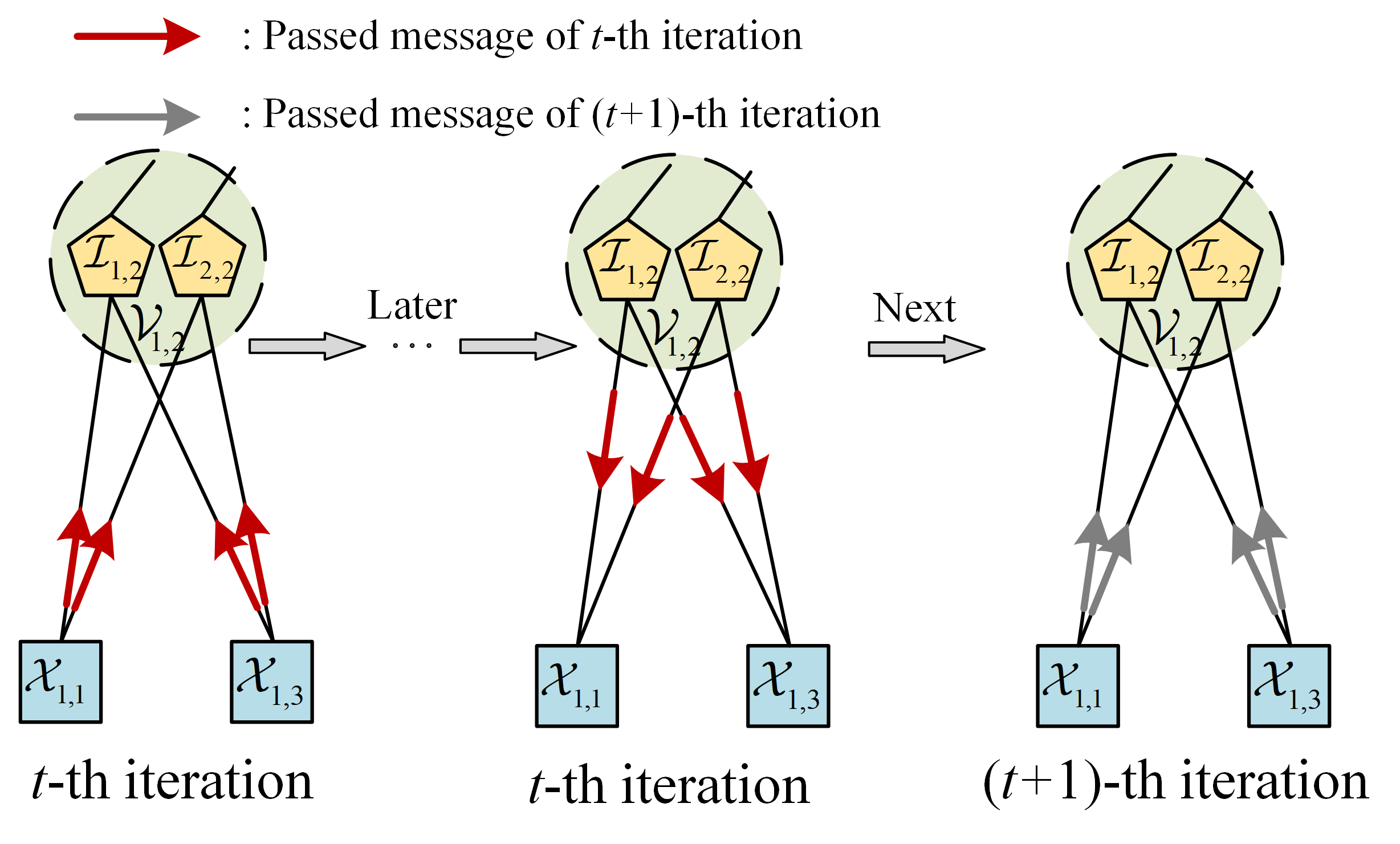}}
	\label{fig9-b}
	\caption{Partial message passing on the factor graph.}
	\label{fig9}
\end{figure}

The partial message passing for NSD-JIDD and ISD-JIDD are described in Fig. \ref{fig9}(a) and Fig. \ref{fig9}(b), respectively, where the same parameters as Fig. \ref{fig5} are considered. For NSD-JIDD, the messages passed by each RN are stored for the UN update in the next iteration. Whereas for ISD-JIDD, it can be interpreted that each UN integrates with \emph{R} INs, instead of being stored for previous messages, which takes advantage of the updated messages promptly.

The overall steps of the ISD-JIDD algorithm are identical to the NSD-JIDD algorithm, only with the improvement of SCMA detection and polar decoding. Specifically, if we replace step 3 in Algorithm 1 by (\ref{39}) and (\ref{11}) to update the UNs and calculate soft messages, respectively, and step 11 performs the L-NB-SCL decoding in Algorithm 2, then the procedures complete the ISD-JIDD algorithm.

\section{Results and Discussions}
\label{sec:5}
Simulation results of the proposed NB-PC-SCMA system over additive white gaussian noise (AWGN) and Rayleigh fading channels are given and analyzed in this section. Particularly, the BER performance, computational complexity and latency are characterized in Sections \ref{sec:5-1} to \ref{sec:5-3}, respectively.
\subsection{BER Performance}
\label{sec:5-1}
In this section, the error performance of the proposed NB-PC-SCMA system is evaluated. The SCMA codebook used in the simulation is designed according to \cite{6966170}. Here, we define the \emph{M}-dimension SCMA codebook with \emph{J} users and \emph{K} resources as $\left( {J,K,M} \right)$. For the receiver, the number of inner iterations for max-log-MPA detection is set to 1. NSD-JIDD and ISD-JIDD algorithms with 5 outer iterations are employed for multi-user detection. The NB-PCs transmitted over AWGN and Rayleigh fading channels are constructed by the Monte Carlo method at ${{{E_b}} \mathord{\left/{\vphantom {{{E_b}} {{N_0}}}} \right.\kern-\nulldelimiterspace} {{N_0}}} = 2$ dB and ${{{E_b}} \mathord{\left/{\vphantom {{{E_b}} {{N_0}}}} \right.\kern-\nulldelimiterspace} {{N_0}}} = 4$ dB, respectively, where the kernel parameter $\gamma$ is set according to \cite{8625284}. Unless otherwise specified, the field order \emph{q} of the NB-PCs is 16. In addition, 16-CRC and 24-CRC are employed for the NB-PCs with $N = 256$ and $N = 1024$, respectively. To limit the computational complexity, the list size \emph{l} is set to 8.

\begin{figure}[!t]
	\centering
	\subfloat[AWGN channels.]{\includegraphics[width=3in]{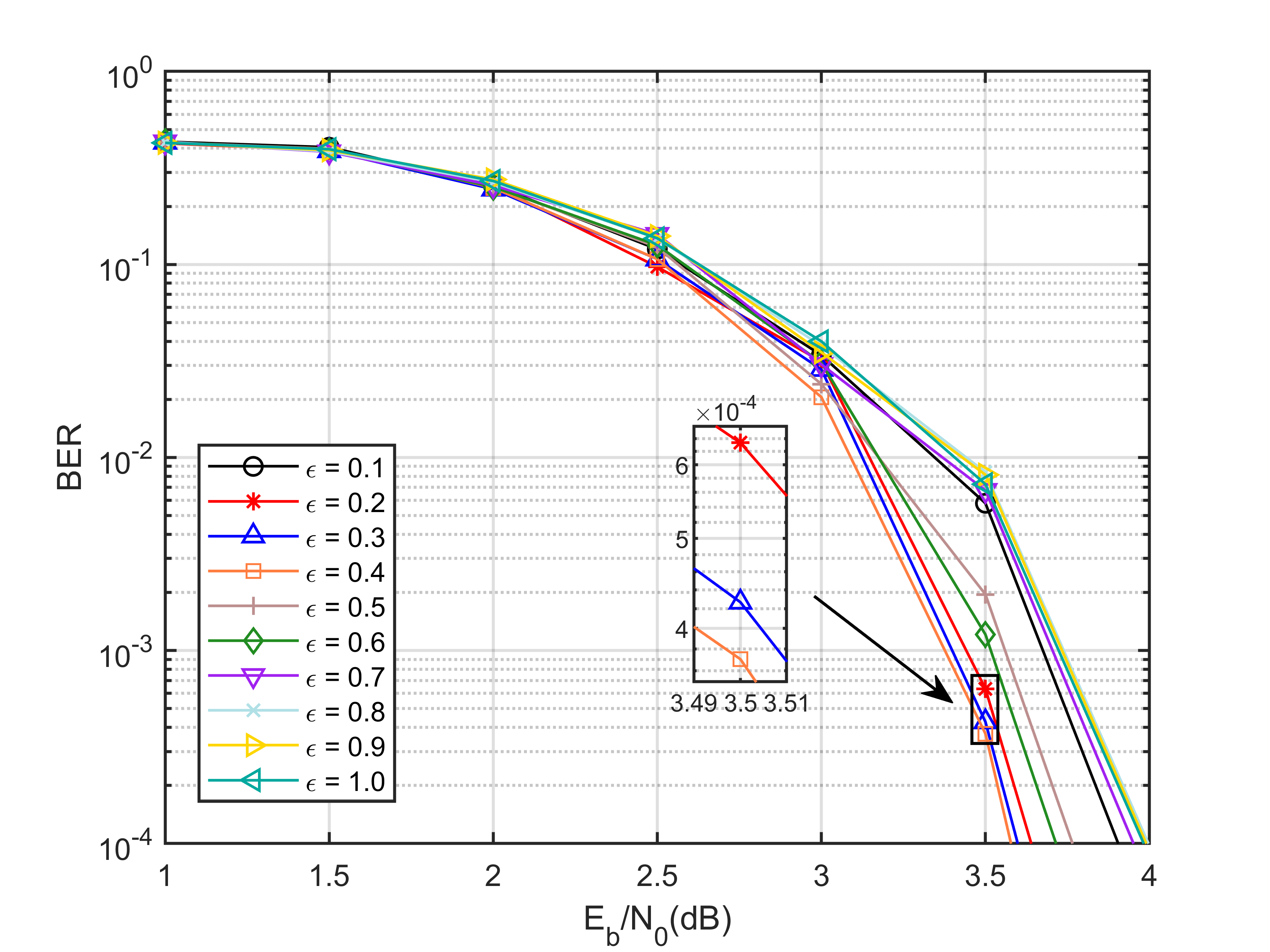}}
	\label{fig10-a}
	\subfloat[Rayleigh fading channels.]{\includegraphics[width=3in]{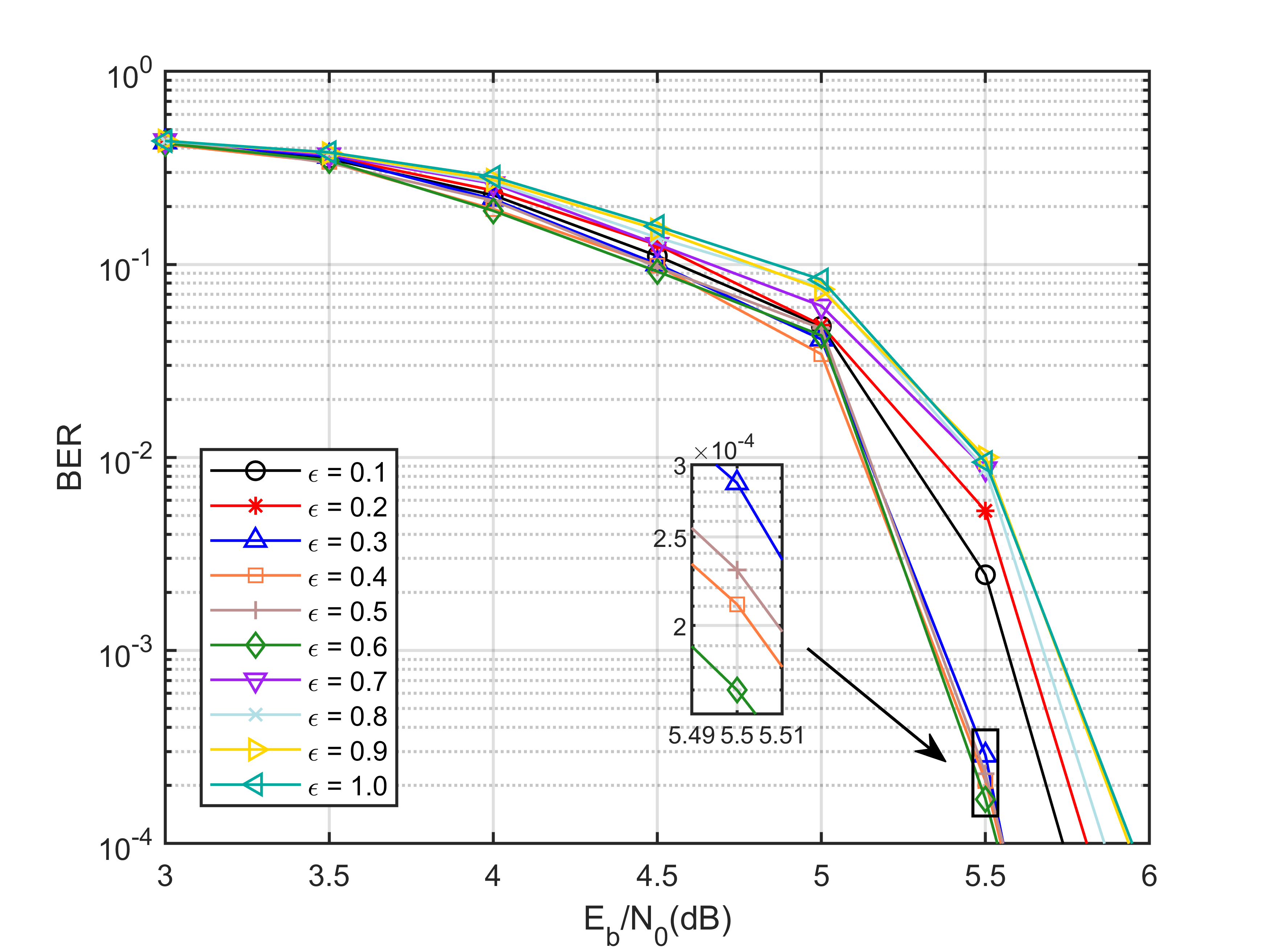}}
	\label{fig10-b}
	\caption{BER performance with different $\varepsilon $, where $N = 256$ and ${R_c} = {1 \mathord{\left/{\vphantom {1 2}} \right.\kern-\nulldelimiterspace} 2}$.}
	\label{fig10}
\end{figure}

Fig. \ref{fig10}(a) and Fig. \ref{fig10}(b) characterize the BER performance of NSD-JIDD with different $\varepsilon $ over AWGN and Rayleigh fading channels, respectively, where the codebook $\left( {6,4,4} \right)$ is considered. It can be seen that $\varepsilon  = 0.4$ and $\varepsilon  = 0.6$ achieve the best performance over AWGN and Rayleigh fading channels, respectively, and hence they are adopted for the following simulations.
\begin{figure}[!t]
	\centering
	\includegraphics[width=3in]{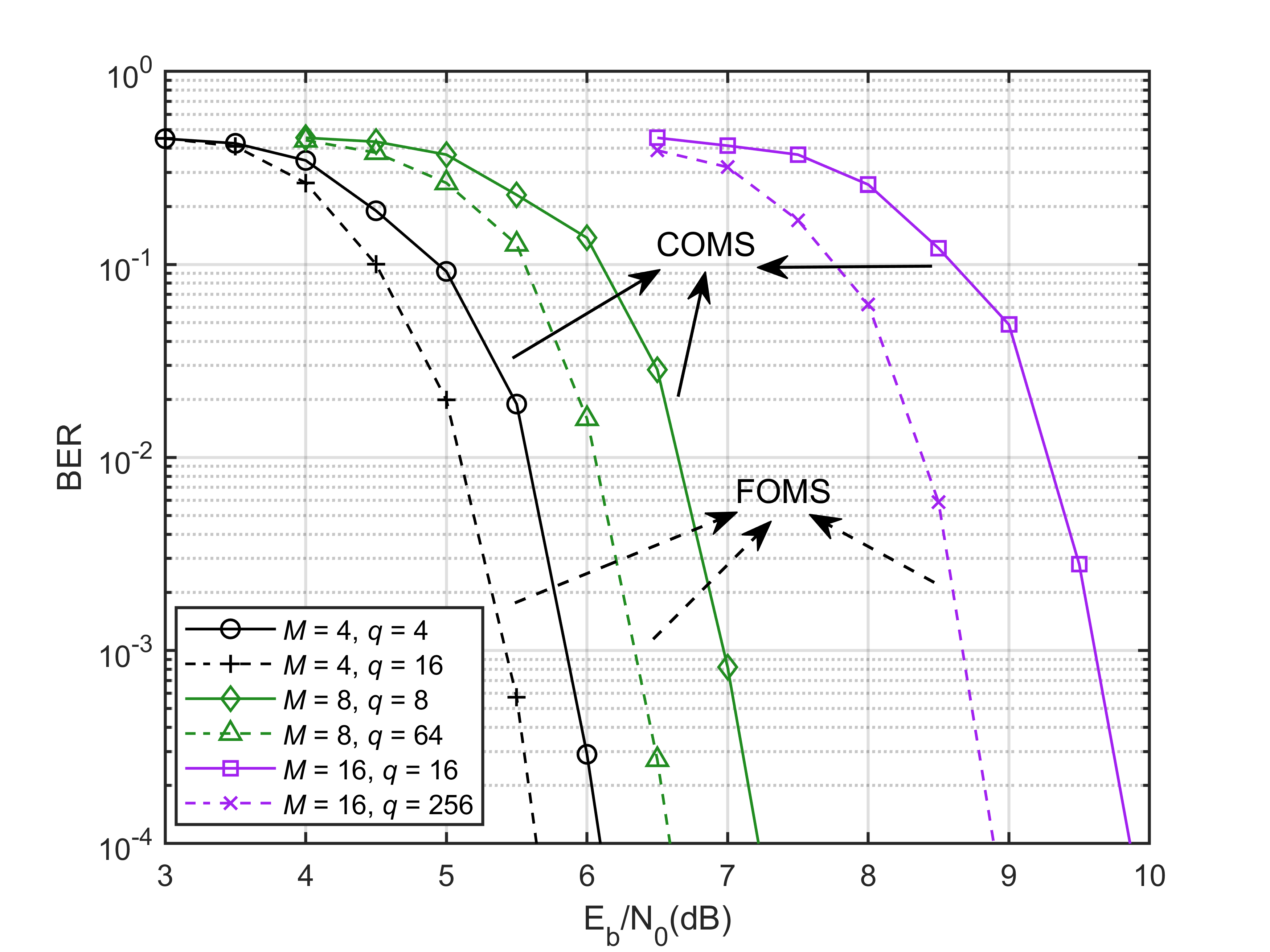}		
	\caption{The BER of NB-PC-SCMA systems with COMS and FOMS, where $J = 6$, $K = 4$, $N = 256$ and ${R_c} = {1 \mathord{\left/{\vphantom {1 2}} \right.\kern-\nulldelimiterspace} 2}$.}
	\label{fig11}
\end{figure}

Fig. \ref{fig11} shows the BER comparison between FOMS and COMS over Rayleigh fading channels, where 4-point, 8-point, and 16-point SCMA modulation are considered. Here, the throughput in terms of bits per symbol \cite{9399239} is defined as $\lambda {R_c}{\log _2}M$. It can be observed that the BER performance of both COMS and FOMS degrades with the increase of throughput. Particularly, FOMS obtains 0.46 dB, 0.64 dB, and 0.97 dB gains versus COMS at the BER of ${10^{ - 4}}$ when bits per symbol are 1.5, 2.25, and 3, respectively. When $q = 16$, FOMS at $M = 4$ outperforms COMS at $M = 16$ about 4.23 dB at the cost of reduced throughput. Thus, we can adjust the order matching to trade-off the throughput and reliability, enabling the NB-PC-SCMA system with flexibility.
\begin{figure}[!t]
	\centering
	\includegraphics[width=3in]{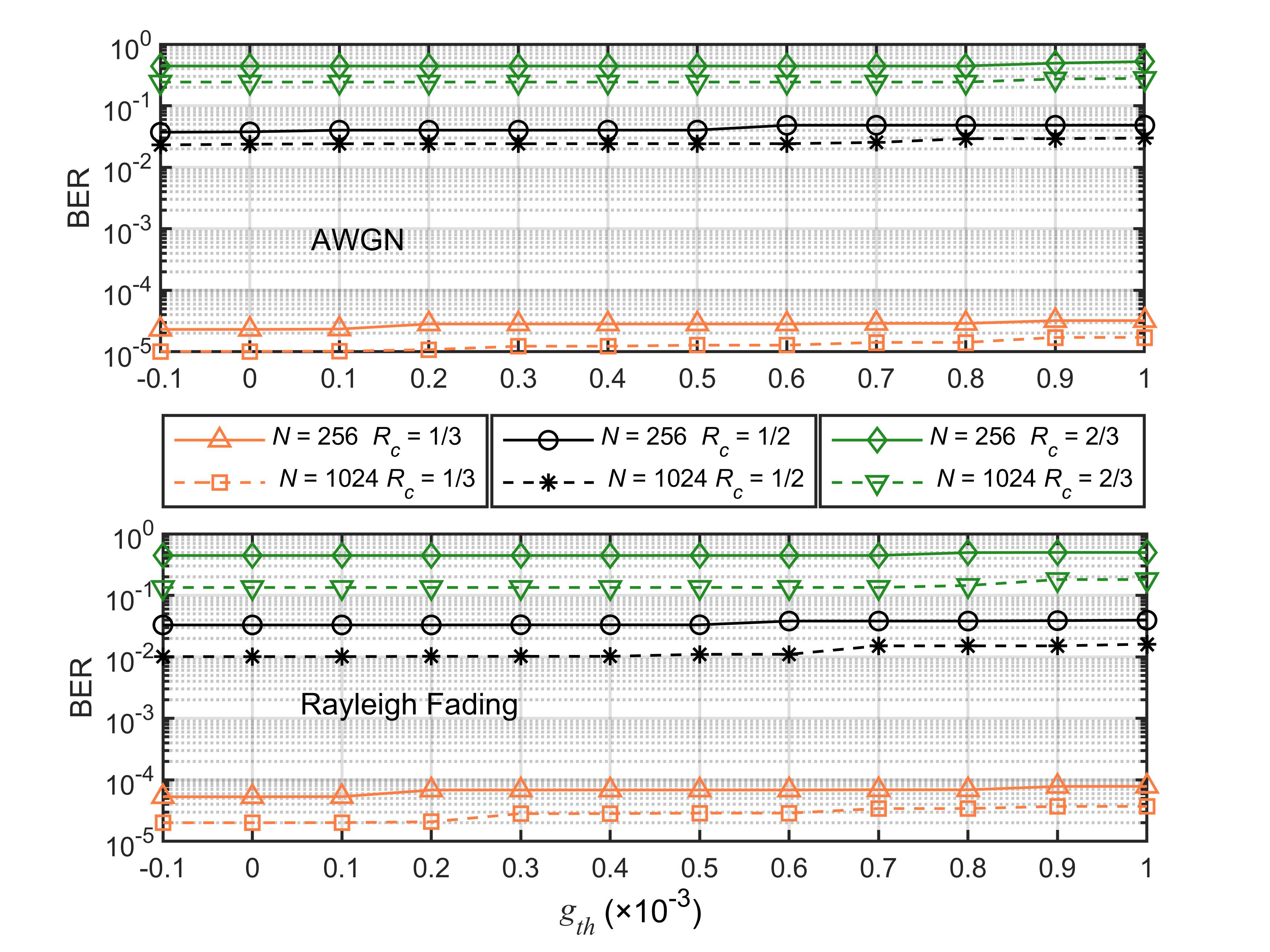}		
	\caption{The relationship between the threshold ${g_{th}}$ and BER of the ISD-JIDD algorithm with different \emph{N} and ${R_c}$, where ${{{E_b}} \mathord{\left/{\vphantom {{{E_b}} {{N_0}}}} \right.
	\kern-\nulldelimiterspace} {{N_0}}}$ is set to 3 dB and 5 dB for AWGN and Rayleigh fading channels when $N = 256$, respectively and where ${{{E_b}} \mathord{\left/{\vphantom {{{E_b}} {{N_0}}}} \right.\kern-\nulldelimiterspace} {{N_0}}}$ is set to 2.5 dB and 4.3 dB for AWGN and Rayleigh fading channels when $N = 1024$, respectively.}
	\label{fig12}
\end{figure}

Before quantifying the performance of the ISD-JIDD algorithm for NB-PC-SCMA systems, we investigate the effect of ${g_{th}}$ in L-NB-SCL decoding on the BER performance in Fig. \ref{fig12}, where the codebook $\left( {6,4,4} \right)$ is considered. Note that ${g_{th}} =  - {10^{ - 4}}$ represents the case when NB-SCL decoding is employed. We can see that within the limit of the threshold, especially at high ${R_c}$, there is no performance degradation when L-NB-SCL decoding is adopted at the receiver. Based on Fig. \ref{fig12}, we select the optimal ${g_{th}}$ for different configurations and calculate the corresponding $\beta $ from  Monte Carlo simulation results, as shown in Table \ref{table3}.
\begin{figure}[!t]
	\centering
	\subfloat[BER performance.]{\includegraphics[width=3in]{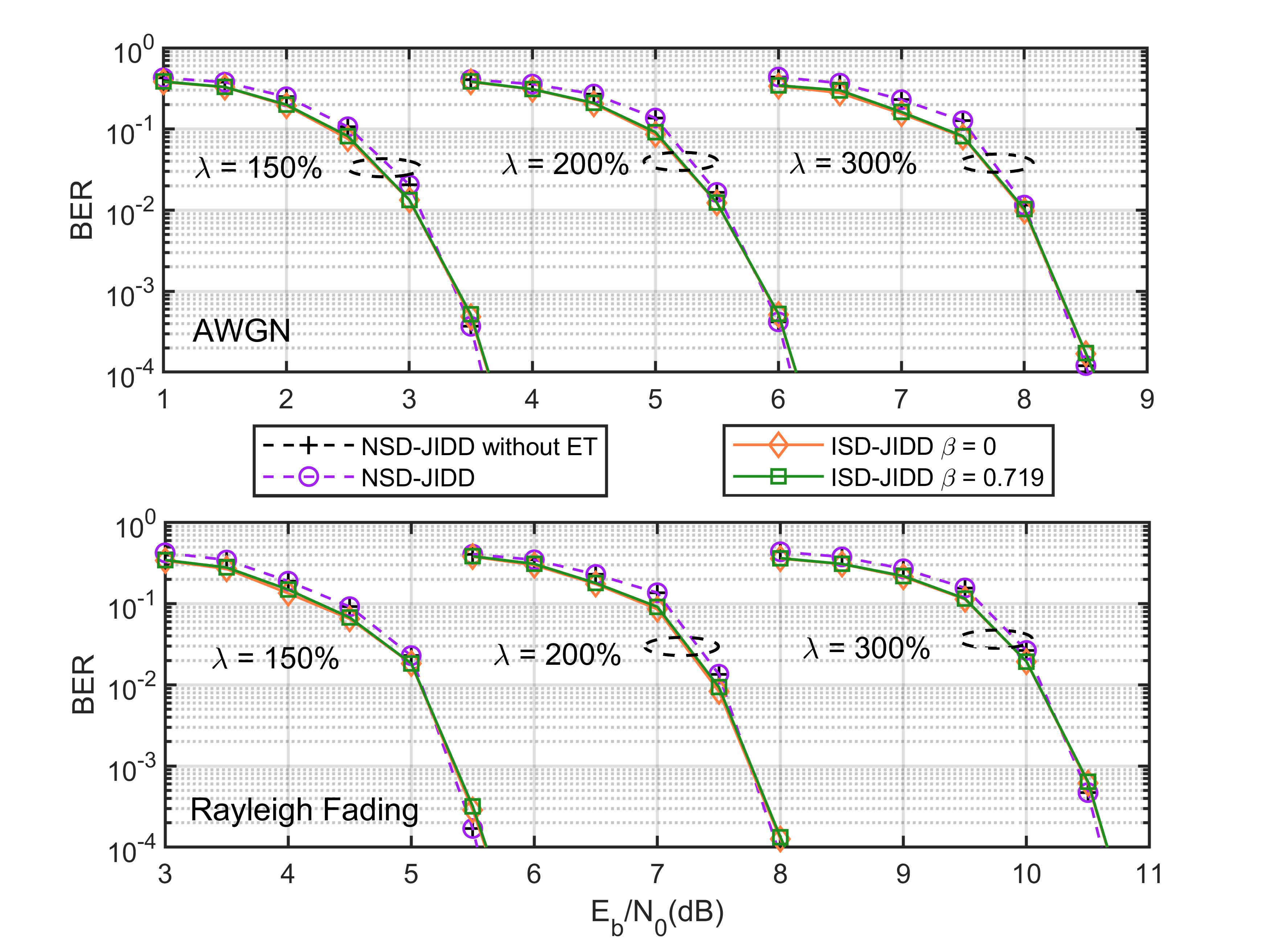}}
	\label{fig13-a}
	\subfloat[convergence performance.]{\includegraphics[width=3in]{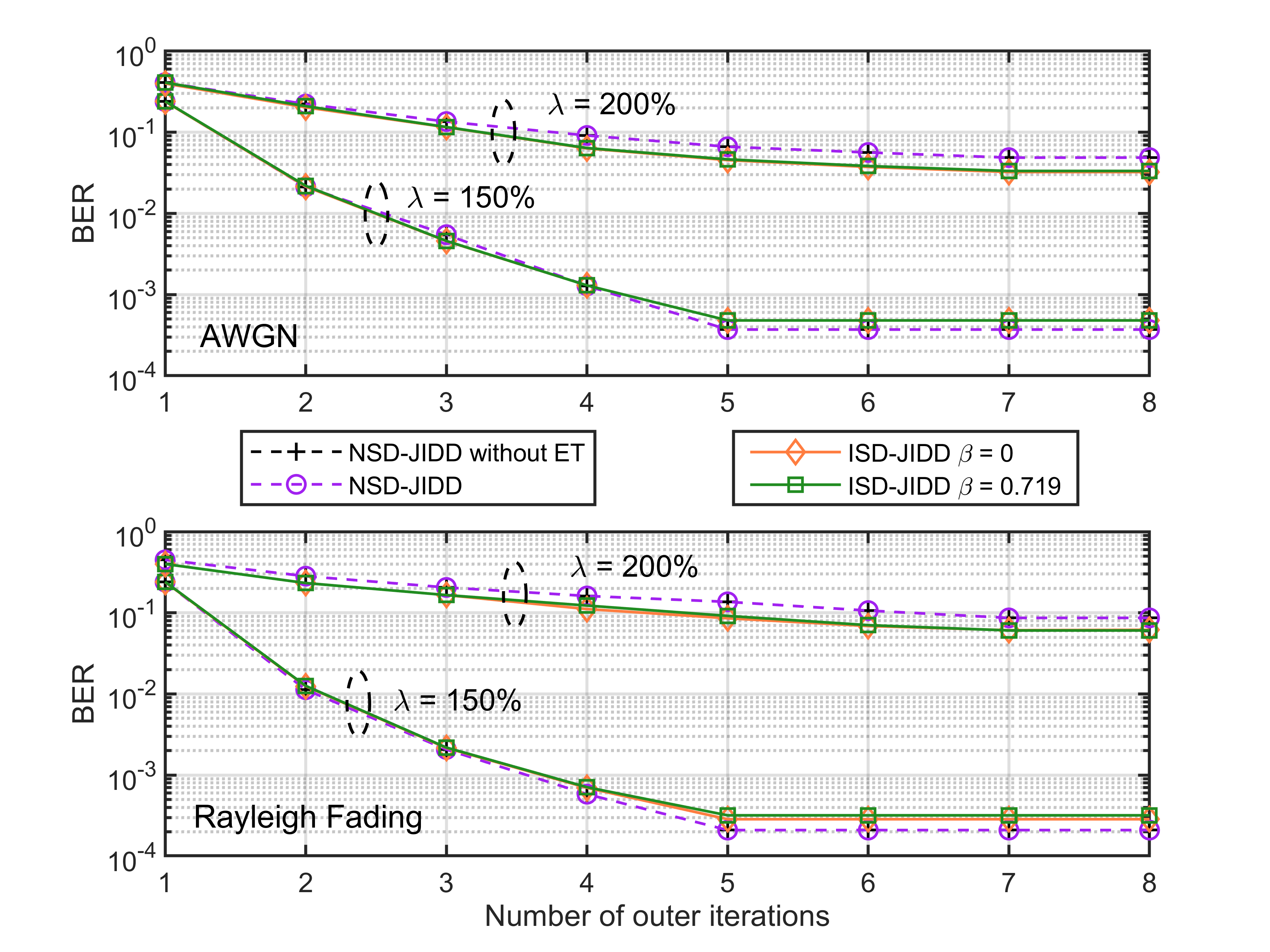}}
	\label{fig13-b}
	\caption{Performance comparison of ISD-JIDD and NSD-JIDD algorithms with different overloads, where $N = 256$ and ${R_c} = {1 \mathord{\left/{\vphantom {1 2}} \right.\kern-\nulldelimiterspace} 2}$}
	\label{fig13}
\end{figure}
\begin{table}[!t]
	\centering
	\renewcommand{\arraystretch}{1.3}
	\caption{Simulation parameters of ISD-JIDD algorithm for different configurations.}
	\label{table3}
	\begin{tabular}{@{}cccccccc@{}}
		\toprule[1pt]
		\multirow{2}{*}{Channel}           & \multirow{2}{*}{Parameters} & \multicolumn{3}{c}{$N = 256$} & \multicolumn{3}{c}{$N = 1024$} \\
		\cmidrule(lr){3-3}\cmidrule(lr){4-4}\cmidrule(lr){5-5}\cmidrule(lr){6-6}\cmidrule(lr){7-7}\cmidrule(l){8-8} 
		&                            & $\frac{1}{3}$   & $\frac{1}{2}$  & $\frac{2}{3}$  & $\frac{1}{3}$   & $\frac{1}{2}$  & $\frac{2}{3}$  \\
		\cmidrule(r){1-1}\cmidrule(lr){2-2}\cmidrule(lr){3-5}\cmidrule(l){6-8}
		\multirow{2}{*}{AWGN}              & ${g_{th}}$                        & 0.1     & 0.5    & 0.8    & 0.2     & 0.7    & 0.8    \\
		& $\beta$                       & 0.714   & 0.719  & 0.698  & 0.871   & 0.906  & 0.871  \\
		\midrule[0.7pt]
		\multirow{2}{*}{\makecell[c]{Rayleigh\\Fading}} & ${g_{th}}$                       & 0.1     & 0.5    & 0.7    & 0.2     & 0.6    & 0.8    \\
		& $\beta$                       & 0.714   & 0.719  & 0.674  & 0.859   & 0.898  & 0.871  \\
		\bottomrule[1pt]
	\end{tabular}
\end{table}
\begin{figure}[!t]
	\centering
	\subfloat[AWGN channels.]{\includegraphics[width=3in]{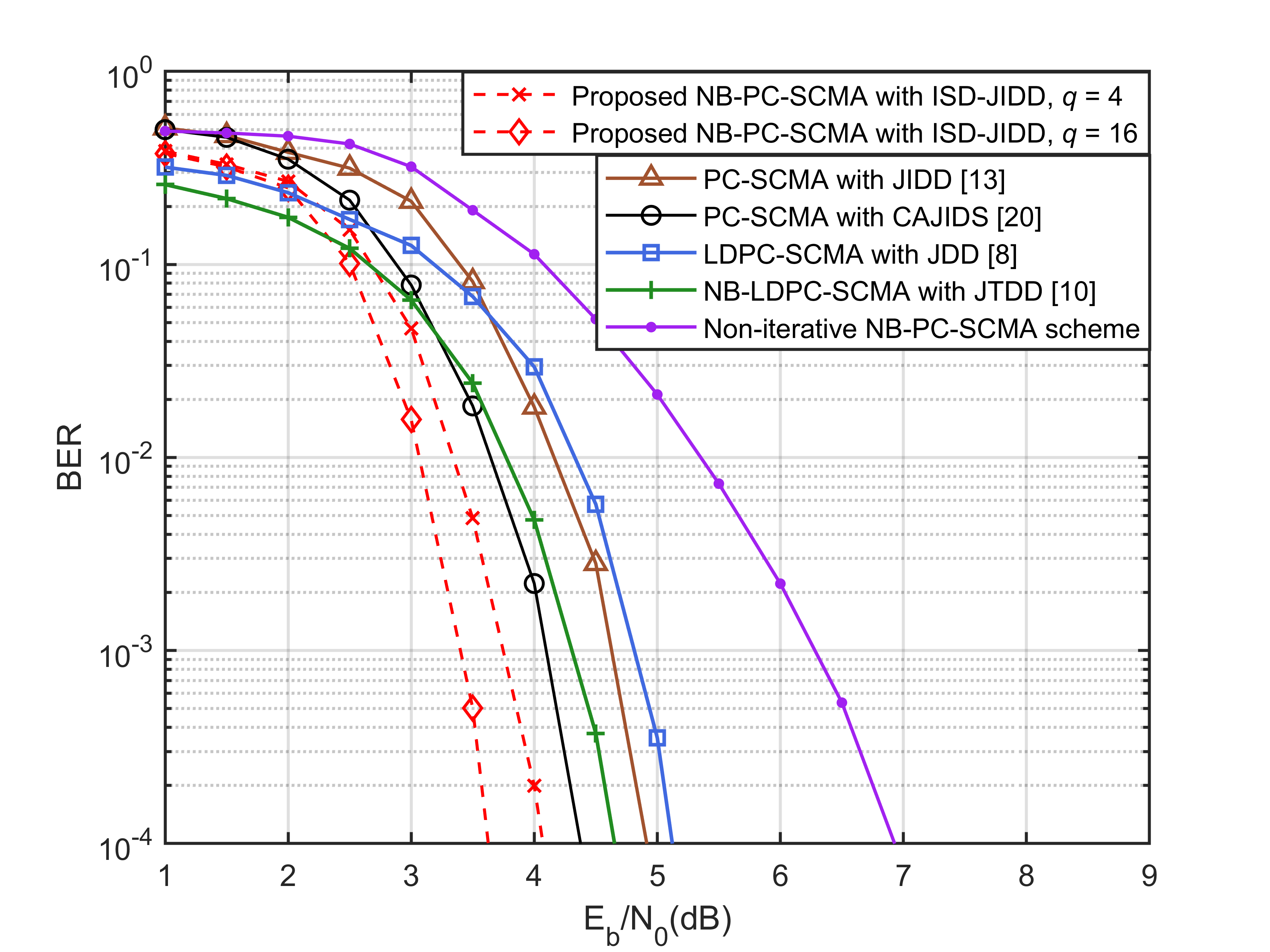}}
	\label{fig14-a}
	\subfloat[Rayleigh fading channels.]{\includegraphics[width=3in]{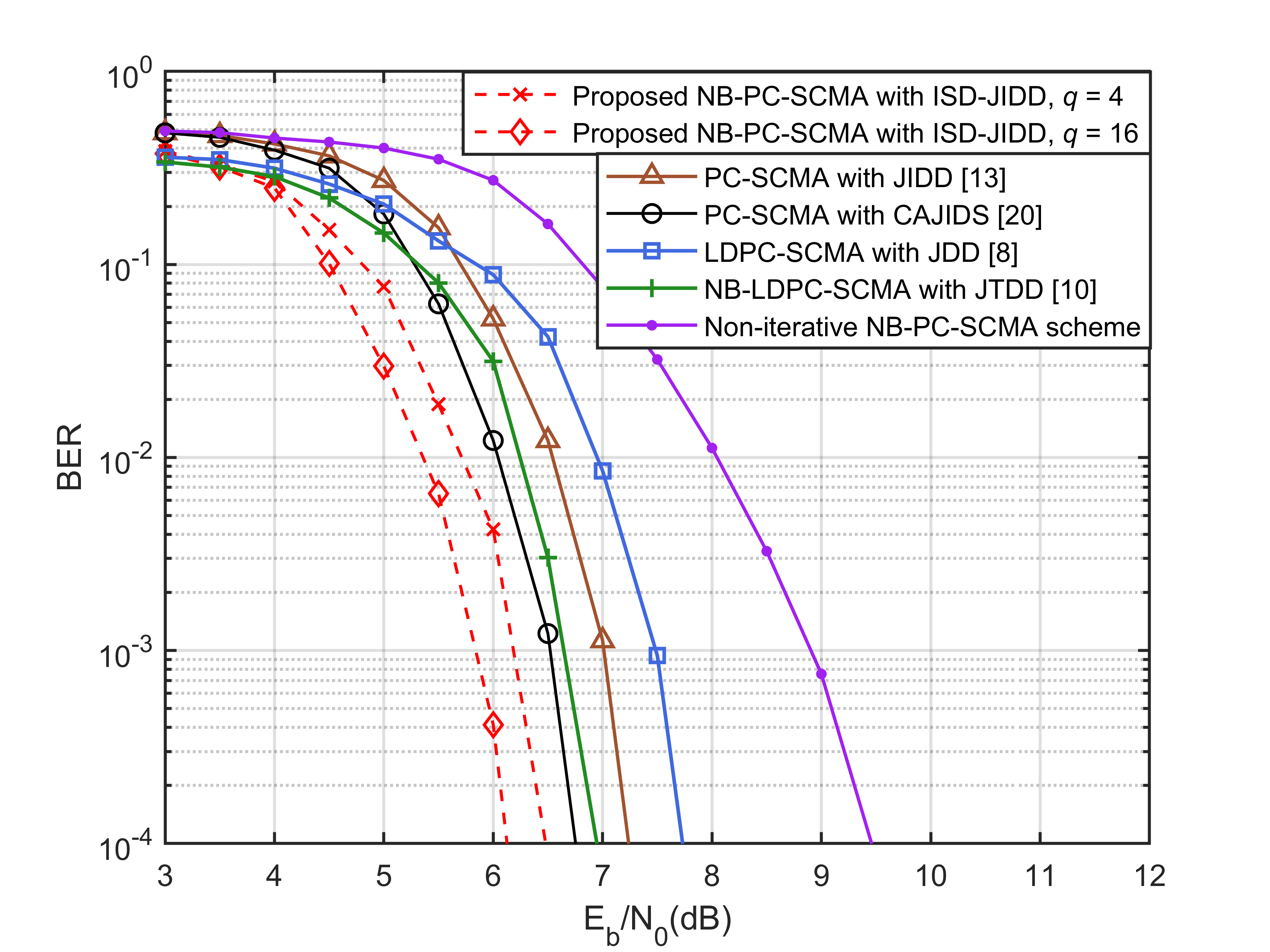}}
	\label{fig14-b}
	\caption{BER comparison of the proposed NB-PC-SCMA to other state-of-the-art coded SCMA schemes, where $N = 256$ and ${R_c} = {1 \mathord{\left/{\vphantom {1 2}} \right.\kern-\nulldelimiterspace} 2}$.}
	\label{fig14}
\end{figure}

The performance comparison of ISD-JIDD and NSD-JIDD with different overloads in terms of BER and convergence are given in Fig. \ref{fig13}(a) and (b), respectively. For $\lambda  = 150\% $, $\lambda  = 200\% $, and $\lambda  = 300\% $, the user number \emph{J} is 6, 12, and 24, respectively, where the corresponding resource number \emph{K} is 4, 6, and 8, respectively. The modulation order \emph{M} is set to 4. Here, the NSD-JIDD without ET and the ISD-JIDD with $\beta  = 0$ are also plotted as the benchmark. 

As seen in Fig. \ref{fig13}(a), a higher user overload $\lambda $ always leads to a poorer BER performance over both AWGN and Rayleigh fading channels despite an improved throughput. ET has no influence on the detection results at all. Moreover, ISD-JIDD shows a better performance than NSD-JIDD at low ${{{E_b}} \mathord{\left/{\vphantom {{{E_b}} {{N_0}}}} \right.\kern-\nulldelimiterspace} {{N_0}}}$ since ISD-JIDD avoids convergence errors with the aid of modified MPA. Overall, ISD-JIDD lags NSD-JIDD only about 0.06 to 0.09 dB at the BER of ${10^{ - 4}}$. Fig. \ref{fig13}(b) shows that both NSD-JIDD and ISD-JIDD require two extra iterations to converge when $\lambda  = 200\% $. Apparently, the convergence performance of ISD-JIDD is not suffered, which is attributed to the ET and damping mechanisms.
\begin{table*}[!t]
	\centering
	\renewcommand{\arraystretch}{1.3}
	\caption{The computational complexity of different PC-SCMA schemes.}
	\label{table5}
	\begin{threeparttable}
		\begin{tabular}{m{0.5cm}<{\centering}!{\vrule width0.9pt}m{3.8cm}<{\centering}|m{3.8cm}<{\centering}!{\vrule 
					width0.9pt}m{3.8cm}<{\centering}|m{3.8cm}<{\centering}}
			\Xhline{1.2pt}
			\multirow{2}{*}{}& \multicolumn{2}{c!{\vrule width0.9pt}}{Proposed NB-PC-SCMA$^{\dagger}$} & \multicolumn{2}{c}{PC-SCMA} \\ \cline{2-5}
			& NSD-JIDD & ISD-JIDD & JIDD \cite{8463448} & CAJIDS \cite{9285274} \\ \hline\hline
			\rule{0pt}{12pt}
			ADD
			\rule{0pt}{12pt}& $E\left[ {2K{d_r}^2{M^{{d_r}}} + } \right.JM({d_u}^2 + \left. {{d_u} - 1)} \right] + J\left[\left( {2q + 1} \right)lN'{\log _2}N' + 2{{{\cal T}_{p_1}}\!\!^{\ddagger}}\right]$ & $E\left[ 2K{d_r}^2{M^{{d_r}}}\right. + \left.JM\left( {{d_u} - 1} \right) \right] + J\left[\left( {2q + 1} \right)lN'{\log _2}N' + 2{{\cal T}_{p_2}}\!\!^{\ddagger}\right]$ & $EK{d_r}\left[ {{M^{{d_r}}}\left( {{d_r} + 2} \right) - 1} \right] + 2JN{\log _2}N$ & $EK{d_r}\left[ {{M^{{d_r}}}\left( {{d_r} + 2} \right) - 1} \right] + J\left(\frac{1}{2}lN{\log _2}N + {{\cal T}_{p_0}}\!\!^{\ddagger}\right)$ \\\hline
			\rule{0pt}{12pt}
			MUL 
			\rule{0pt}{12pt}& $EK{d_r}{M^{{d_r}}}\left( {{d_r} + 3} \right)$ & $EK{d_r}{M^{{d_r}}}\left( {{d_r} + 3} \right)$ & $E\left\{ K{d_r}\left[ M\left( {{d_u} - 1} \right) \right.\right.+ {M^{{d_r}}}\cdot \left.\left. \left( {2{d_r} + 4} \right) \right] + JM\left( {{d_u} - 1} \right) \right\} + 4JN{\log _2}N$ & $E\left\{ K{d_r}\left[ M\left( {{d_u} - 1} \right) \right.\right. + {M^{{d_r}}}\cdot\left.\left.\left( {2{d_r} + 4} \right) \right] + JM\left( {{d_u} - 1} \right) \right\} + JlN{\log _2}N$ \\\hline
			\rule{0pt}{12pt}
			CMP
			\rule{0pt}{12pt} & $EK{d_r}M\left( {{M^{{d_r} - 1}} + M - 2} \right) + J\left[2\left( {q - 1} \right)lN'{\log _2}N'\right. + \left.(q - 1){{\cal T}_{p_1}}\right]$ & $EK{d_r}M\left( {{M^{{d_r} - 1}} - 1} \right) + J\left[2\left( {q - 1} \right)lN'{\log _2}N'\right. + \left.(q - 1){{\cal T}_{p_2}}\right]$ & $6JN{\log _2}N$ & $J\left(2lN{\log _2}N + {{\cal T}_{p_0}}\right)$\\\hline
			\rule{0pt}{0.5pt}
			XOR  
			\rule{0pt}{0.5pt} & $\frac{1}{2}JlN'{\log _2}N'$ & $\frac{1}{2}JlN'{\log _2}N'$ & 0 & $\frac{1}{2}JlN{\log _2}N$ \\\hline
			\rule{0pt}{0.5pt}
			EXP
			\rule{0pt}{0.5pt} & 0 & 0 & $EK{d_r}{M^{{d_r}}}$ & $EK{d_r}{M^{{d_r}}}$ \\
			\Xhline{1.2pt}
		\end{tabular}
		\begin{tablenotes}
			\footnotesize
			\item[$\dagger$] The element-wise MUL of non-binary polar decoders are evaluated as ADD since such operations are represented as the addition of exponents.
			\item[$\ddagger$] ${\cal T}_{p_0}$, ${\cal T}_{p_1}$, and ${\cal T}_{p_2}$ denote the number of search paths in the SCL, NB-SCL, and L-NB-SCL decoding, respectively.
		\end{tablenotes}
	\end{threeparttable}
\end{table*}

Finally, we compare the BER performance of the proposed NB-PC-SCMA with ISD-JIDD to other state-of-the-art schemes, where the codebook $\left( {6,4,4} \right)$ is considered. To be more specific, LDPC-SCMA with JDD \cite{7848813}, non-binary LDPC (NB-LDPC) coded SCMA (NB-LDPC-SCMA) with joint trellis based joint decoding and detection (JTDD) \cite{8344383}, PC-SCMA with JIDD \cite{8463448}, and PC-SCMA with CRC aided joint iterative detection and SCL decoding (CAJIDS) \cite{9285274} are simulated as counterparts. Especially, the PC-SCMA schemes with JIDD or CAJIDS can also be considered as benchmarks for different left-to-right strategies during polar decoding. The LDPC and NB-LDPC codes in the simulations are constructed in \cite{4432803} and \cite{4155118}, respectively, where the min-sum and extended min-sum decoders with 30 inner iterations are employed in the receiver, respectively. In addition, the non-iterative NB-PC-SCMA scheme is included to analyze the iterative gain, where the result is based on the traditional COMS.

As demonstrated in Fig. \ref{fig14}(a), the NB-PC-SCMA system with ISD-JIDD outperforms PC-SCMA with CAJIDS and PC-SCMA with JIDD about 0.32 dB and 0.86 dB at the BER of ${10^{ - 4}}$ over AWGN channels when $q = 4$. If the field order \emph{q} is set to 16 with the aid of FOMS, the performance gains increase to 0.80 dB and 1.34 dB, respectively. Explicitly, an iterative gain of at least 2.84 dB can be observed when compared to the non-iterative counterpart. As for another non-binary coding scheme, i.e., NB-LDPC-SCMA, a 0.59 dB performance gain is attained by NB-PC-SCMA with $q = 4$. In contrast to the COMS-based NB-LDPC-SCMA in \cite{8344383}, our proposed FOMS-based NB-PC-SCMA achieves up to 1.07 dB gain when $q = 16$. 

For Rayleigh fading channels, the BER performance of all schemes is inferior to that over AWGN channels, while similar comparison results can be observed in Fig. \ref{fig14}(b). To be specific, the iterative gain of the ISD-JIDD achieves at least 2.98 dB. When $q = 16$, the NB-PC-SCMA system with ISD-JIDD obtains 0.67 dB and 1.18 dB gain compared to the binary counterparts with CAJIDS and JIDD, respectively. Furthermore, our proposed FOMS-based NB-PC-SCMA with $q = 16$ outperforms the COMS-based NB-LDPC-SCMA about 0.83 dB. Overall, we can find that the proposed NB-PC-SCMA system with ISD-JIDD always shows the best error correction performance.
\subsection{Complexity}
\label{sec:5-2}
This section evaluates the computational complexity of the receiver for the proposed NB-PC-SCMA scheme, which mainly lies in the SCMA detector and the polar decoder. Here, the computational complexity is quantified by the number of floating-point operations per second per outer iteration. Specifically, addition, multiplication, comparison, exclusive or, and exponential arithmetic operations are counted, which are represented by ADD, MUL, CMP, XOR, and EXP, respectively.

Typically, the complexity of the polar decoder is less than that of the LDPC decoder \cite{7980697}. Thus, other outstanding PC-SCMA schemes, e.g., \cite{8463448} and \cite{9285274}, are considered as comparison benchmarks. The computational complexity of different PC-SCMA schemes with various receiver strategies is presented in Table \ref{table5}. It can be seen that neither NSD-JIDD nor ISD-JIDD includes EXP operations since the max-log-MPA algorithm is applied. As \emph{q} and \emph{N} increase, the exponentially-growing ${\cal T}_{p_1}$ leads to a high computational complexity of NSD-JIDD. By contrast, the ISD-JIDD receiver employs L-NB-SCL decoding to simplify the path search pattern, which yields a much lower ${\cal T}_{p_2}$ than ${\cal T}_{p_1}$. In addition, partial ADD and CMP operations of NSD-JIDD are also reduced by the modified MPA detection in the ISD-JIDD.

\begin{figure}[!t]
	\centering
	\includegraphics[width=3in]{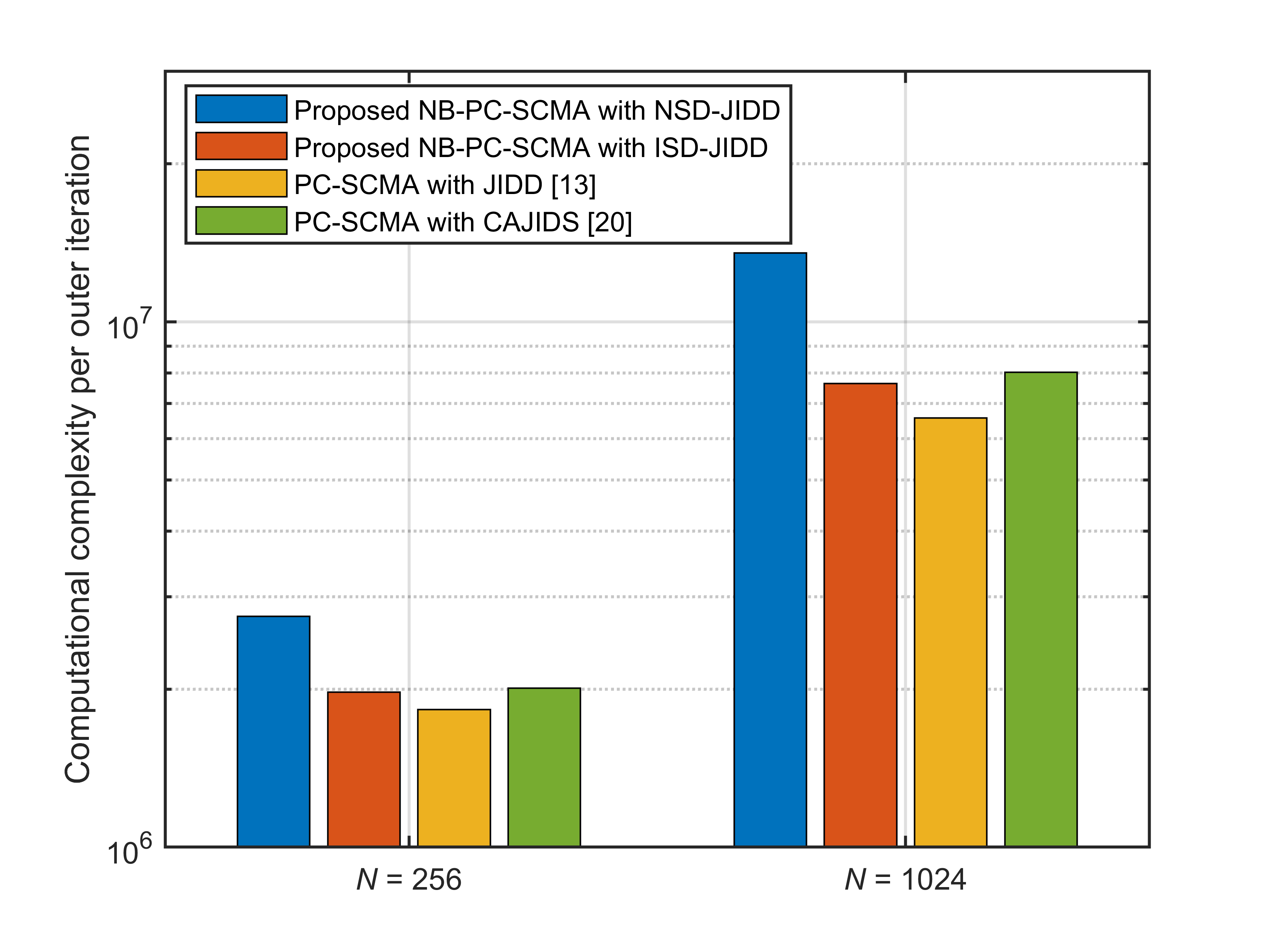}		
	\caption{The computational complexity per outer iteration of different PC-SCMA schemes with codebook $\left( {6,4,4} \right)$, where $q=4$ is considered for NB-PC-SCMA.}
	\label{fig15}
\end{figure}

Fig. \ref{fig15} illustrates the computational complexity per outer iteration of the proposed NB-PC-SCMA system and benchmarks with different code lengths, where the field size $q = 4$ and the codebook $\left( {6,4,4} \right)$ are considered. We can see that when $N = 256$, NSD-JIDD yields a 48\% and 36\% higher complexity over JIDD and CAJIDS, respectively. In comparison, the complexity of ISD-JIDD is 31\% less than that of NSD-JIDD and only increases by 7\% of JIDD. When $N = 1024$, a 39\% complexity reduction compared to NSD-JIDD can be observed.
\subsection{Latency}
\label{sec:5-3}
Here, clock cycles are used to measure the latency gain of the NB-PC-SCMA system compared to the binary counterpart. Suppose that all parallelizable instructions are carried out in one clock cycle. SCMA detection takes only three cycles, i.e., each of which corresponds RN update, UN update, and calculation of soft information. The latency of SCMA detection is negligible compared to polar decoding. Therefore, we will focus on the impact of polar decoding on the receiver latency.

The cycles of the binary SCL decoder can be calculated as ${\Gamma _{SCL}} = 2N - 2 + D$ \cite{7742998} with \emph{D} cycles for path sorting and $2N - 2$ cycles for SC decoding. The latency of the SCAN decoder can be expressed as ${\Gamma _{SCAN}} = {T_{SCAN}}\left( {2N - 2} \right)$, where ${T_{SCAN}}$ is the number of SCAN inner iterations. The latency of the L-NB-SCL decoder is written as ${\Gamma _{L-NB - SCL}} = 2N' - 2 + \left( {1 - \beta } \right)D'$. Thus, the latency for the JIDD in \cite{8463448} can be expressed as
\begin{equation}
	{\Gamma _{JIDD}} = T \cdot {T_{SCAN}}\left( {2N - 2} \right) = T\left( {2N - 2} \right).
	\label{43}
\end{equation}

According to \cite{9285274}, the latency for the CAJIDS can be calculated as
\begin{equation}
	{\Gamma _{CAJIDS}} = \sum\limits_{t = 1}^T {2{N_t} - 2 + {D_t}},
	\label{44}
\end{equation}
where ${N_t}$ and ${D_t}$ denote the number of decoded bits and information bits before ET in the \emph{t}-th iteration for the distributed CRC aided SCL decoding, respectively. The latency for our proposed ISD-JIDD can be expressed as
\begin{equation}
	\begin{aligned}
		{\Gamma _{ISD - JIDD}} &= {T_a}\left( {2N' - 2 + \left( {1 - \beta } \right)D'} \right)\\
		& = \frac{{{T_a}}}{p}\left[ {2N - 2p + \left( {1 - \beta } \right)D} \right],
	\end{aligned}	
	\label{45}
\end{equation}
where ${T_a}$ denotes the average number of iterations for the ISD-JIDD. To make a fair comparison, we employ the latency of a multiuser receiver with conventional SCL decoding as the benchmark. Then, the latency gain for algorithm $\bm{{\rm{X}}}$ can be written as
\begin{equation}
	{{\cal G}_{latency,{\bm{{\rm{X}}}}}} = \frac{{T \cdot {\Gamma _{SCL}} - {\Gamma _{\bm{{\rm{X}}}}}}}{{T \cdot {\Gamma _{SCL}}}},
	\label{46}
\end{equation}
where ${\Gamma _{\bm{{\rm{X}}}}}$ denotes the latency for a given algorithm $\bm{{\rm{X}}}$, i.e., ${\Gamma _{\bm{{\rm{X}}}}} \in \left\{ {{\Gamma _{JIDD}},{\Gamma _{CAJIDS}},{\Gamma _{ISD - JIDD}}} \right\}$.
\begin{figure}[!t]
	\centering
	\includegraphics[width=3in]{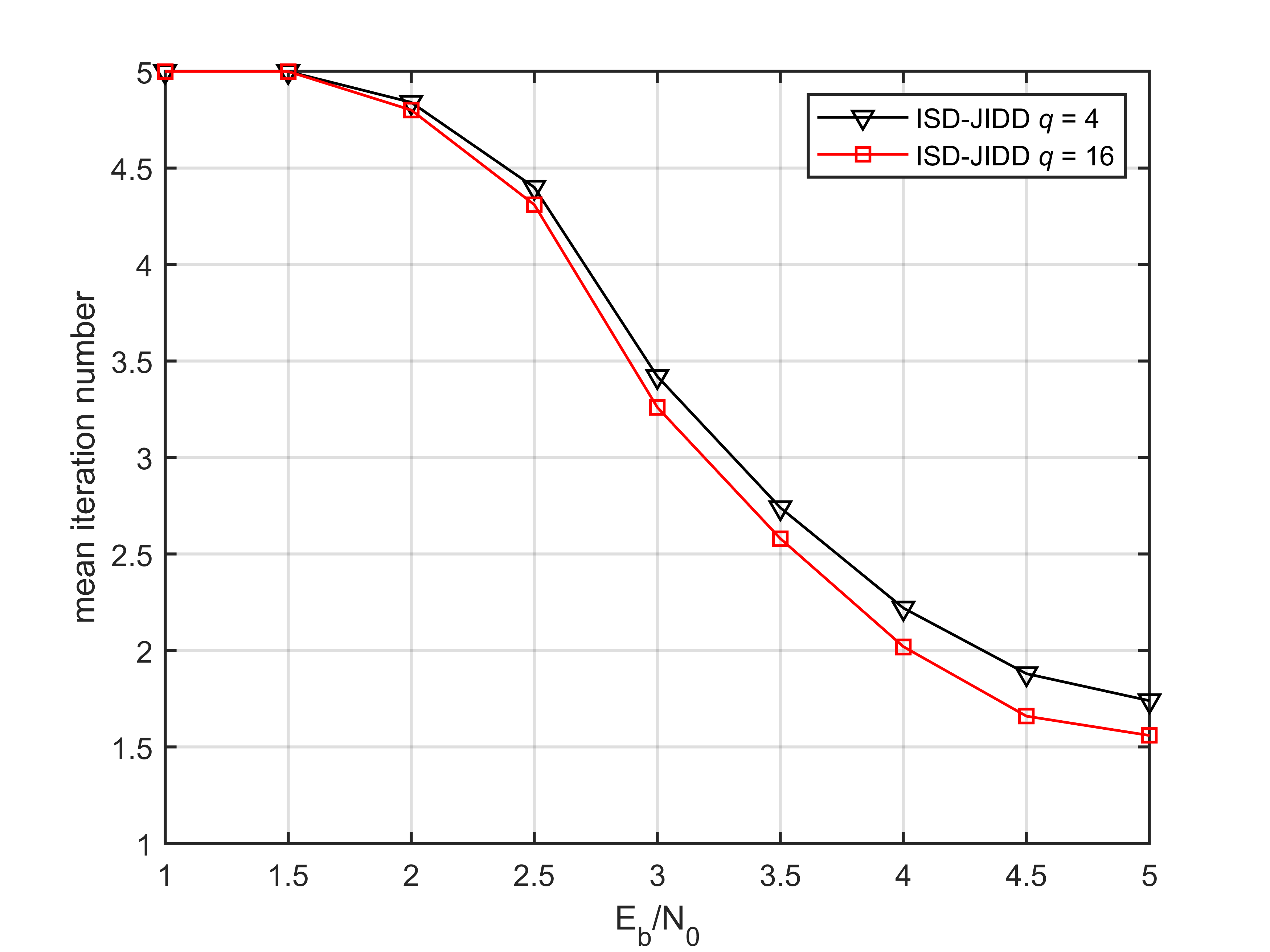}		
	\caption{The average number of iterations for ISD-JIDD over AWGN channels, where $N = 256$ and ${R_c} = {1 \mathord{\left/{\vphantom {1 2}} \right.\kern-\nulldelimiterspace} 2}$.}
	\label{fig16}
\end{figure}
\begin{figure}[!t]
	\centering
	\includegraphics[width=3in]{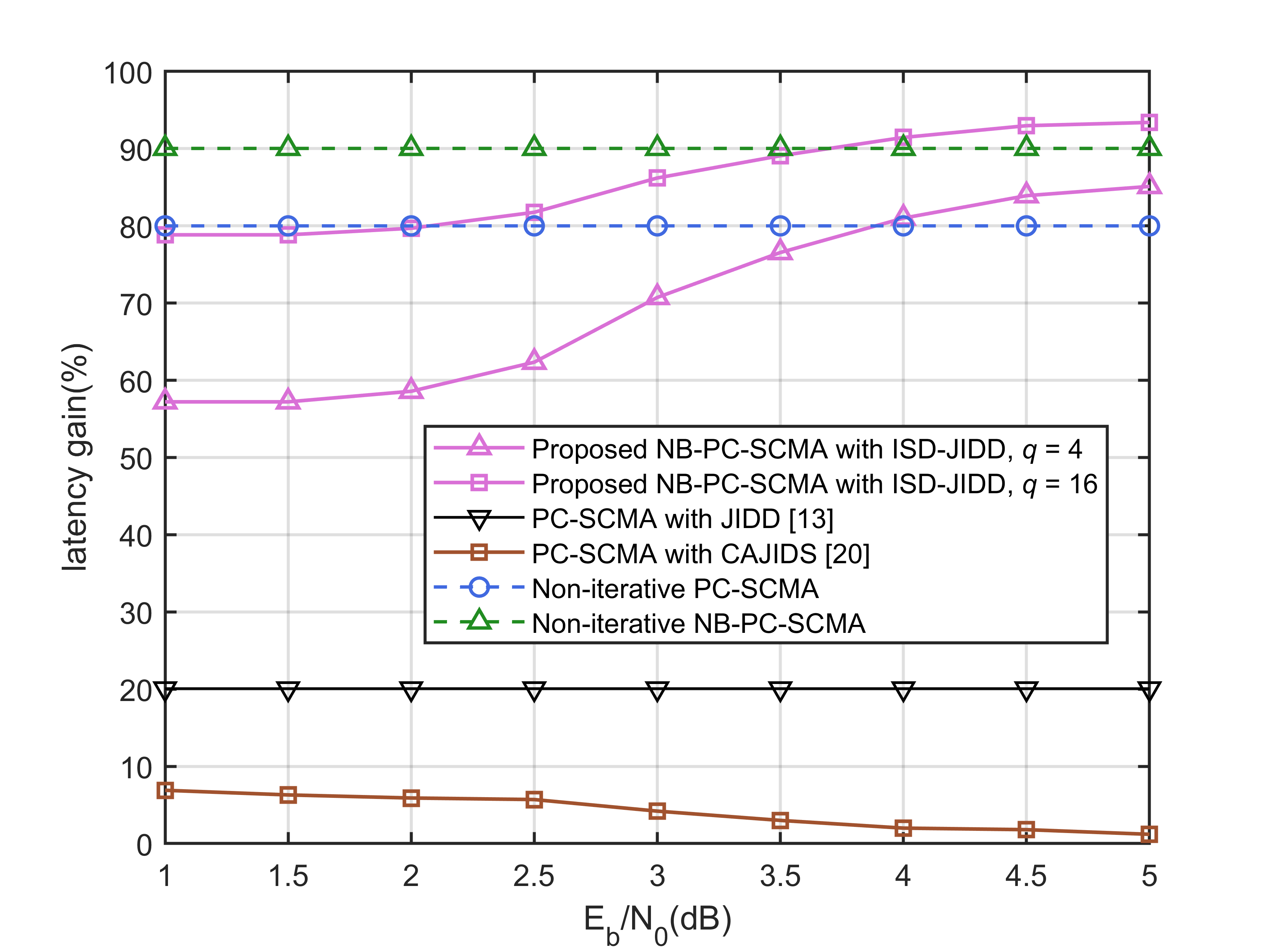}		
	\caption{Comparison of latency gain over AWGN channels, where $N = 256$ and ${R_c} = {1 \mathord{\left/
				{\vphantom {1 2}} \right.\kern-\nulldelimiterspace} 2}$.}
	\label{fig17}
\end{figure}

Before characterizing the latency performance, ${T_a}$ versus ${{{E_b}} \mathord{\left/{\vphantom {{{E_b}} {{N_0}}}} \right.\kern-\nulldelimiterspace} {{N_0}}}$ over AWGN channels is presented in Fig. \ref{fig16}. It can be observed that with the improvement of ${{{E_b}} \mathord{\left/{\vphantom {{{E_b}} {{N_0}}}} \right.\kern-\nulldelimiterspace} {{N_0}}}$, the ET mechanism leads to a decrease of ${T_a}$. Especially at ${{{E_b}} \mathord{\left/{\vphantom {{{E_b}} {{N_0}}}} \right.\kern-\nulldelimiterspace} {{N_0}}} = 5$ dB, the ISD-JIDD with $q = 16$ converges and activates ET after 1.55 average iterations, which contrasts significantly with the maximum number of iterations $T = 5$.

The latency gain performance for different schemes is presented in Fig. \ref{fig17}, where the same parameters as Fig. \ref{fig14} are considered. In particular, both PC-SCMA and NB-PC-SCMA with non-iterative designs are plotted as the baseline for the iteration scheme. It can be found that for the PC-SCMA system, JIDD has a 20\% latency gain regardless of ${{{E_b}} \mathord{\left/{\vphantom {{{E_b}} {{N_0}}}} \right.\kern-\nulldelimiterspace} {{N_0}}}$, while CAJIDS exhibits a weak latency gain at high ${{{E_b}} \mathord{\left/{\vphantom {{{E_b}} {{N_0}}}} \right.\kern-\nulldelimiterspace} {{N_0}}}$. By contrast, for the proposed NB-PC-SCMA system, ISD-JIDD obtains about 57\% to 92\% latency gain, attributed to the characteristics of NB-PC, ET mechanism and L-NB-SCL decoding. 

Observe also from Fig. \ref{fig17} that the non-iterative scheme achieves up to 90\% latency reduction since the receiver does not consider a feedback iteration mechanism. At low ${{{E_b}} \mathord{\left/{\vphantom {{{E_b}} {{N_0}}}} \right.\kern-\nulldelimiterspace} {{N_0}}}$, ISD-JIDD with $q = 4$ takes approximately twice the latency of the non-iterative PC-SCMA. However, the BER performance of the non-iterative system deteriorates significantly as shown in Fig. \ref{fig14}. Benefiting from FOMS, the latency gain of the proposed system becomes larger as \emph{q} increases. Note that when ${{{E_b}} \mathord{\left/{\vphantom {{{E_b}} {{N_0}}}} \right.\kern-\nulldelimiterspace} {{N_0}}} = 5$ dB, a latency saving of 92\% can be observed by ISD-JIDD with $q = 16$, which even outperforms that of non-iterative NB-PC-SCMA systems.
\section{Conclusion}
\label{sec:6}
In this paper, we have presented the design of the FOMS-based NB-PC- SCMA system for the first time. Specifically, an NSD-JIDD receiver have been proposed to gurantee the BER performance. Moreover, the ISD-JIDD algorithm with L-NB-SCL decoding and modified MPA have been proposed to reduce computational complexity and convergence error. Simulation results show that the proposed NB-PC-SCMA system leads to the best BER performance with up to 92\% latency gain compared to other state-of-the-art architecture. Furthermore, ISD-JIDD achieve at leat 31\% complexity reduction over NSD-JIDD without significant BER loss. In our future work, we will extend the proposed NB-PC-SCMA scheme to the scenarios with imperfect channel state information where channel estimation technologies are considered.

\ifCLASSOPTIONcaptionsoff
  \newpage
\fi

\bibliographystyle{IEEEtran}
\bibliography{reference}

\vspace*{-29mm}
\end{document}